\documentclass{ieeeaccess}
\usepackage{cite}
\usepackage{amsmath,amssymb,amsfonts}
\usepackage{graphicx}
\graphicspath{ {images/} }
\usepackage{algorithm}
\usepackage{amsmath,bm,amssymb}
\usepackage{algcompatible}
\usepackage{hyperref}
\usepackage{graphics}

\usepackage{textcomp}
\usepackage{import}
\def\BibTeX{{\rm B\kern-.05em{\sc i\kern-.025em b}\kern-.08em
    T\kern-.1667em\lower.7ex\hbox{E}\kern-.125emX}}

\begin{document}

\history{Date of publication xxxx 00, 0000, date of current version xxxx 00, 0000.}
\doi{XXXXXX/ACCESS.XXXX.DOI}
% \usepackage[english]{bibtex}

% \author[1]{Kazi Nazmul Haque}
% \author[1]{Rajib Rana}
% \author[2,3]{Bj\"orn W.\ Schuller}

% \affil[1]{University of Southern Queensland, Australia}
% \affil[2]{GLAM -- Group on Language, Audio, \& Music, Imperial College London, UK}
% \affil[3]{Chair of Embedded Intelligence for Health Care and Wellbeing, University of Augsburg, Germany}

\title{High-Fidelity Audio Generation and Representation Learning with Guided Adversarial Autoencoder}
% \author{\uppercase{Kazi Nazmul Haque}\authorrefmark{1}, \IEEEmembership{Student Member, IEEE},
% \uppercase{Rajib Rana\authorrefmark{1}, and Bj\"orn W.\ Schuller,
% Jr}.\authorrefmark{2,3},
% \IEEEmembership{Member, IEEE}}

\author{\uppercase{Kazi Nazmul Haque}\authorrefmark{1}, 
\uppercase{Rajib Rana\authorrefmark{1}, and Bj\"orn W.\ Schuller,
Jr}.\authorrefmark{2,3}}

\address[1]{University of Southern Queensland, Australia}
\address[2]{GLAM -- Group on Language, Audio, \& Music, Imperial College London, UK}
\address[3]{Chair of Embedded Intelligence for Health Care and Wellbeing, University of Augsburg, Germany}
% \tfootnote{This paragraph of the first footnote will contain support 
% information, including sponsor and financial support acknowledgment. For 
% example, ``This work was supported in part by the U.S. Department of 
% Commerce under Grant BS123456.''}

\markboth
{Author \headeretal: Preparation of Papers for IEEE TRANSACTIONS and JOURNALS}
{Author \headeretal: Preparation of Papers for IEEE TRANSACTIONS and JOURNALS}

\corresp{Corresponding author: Kazi Nazmul Haque (e-mail: shezan.huq@gmail.com).}

\begin{abstract}
Generating high-fidelity conditional audio samples and learning representation from unlabelled audio data are two challenging problems in machine learning research. Recent advances in the Generative Adversarial Neural Networks (GAN) architectures show great promise in addressing these challenges. To learn powerful representation using GAN architecture, it requires superior sample generation quality, which requires an enormous amount of labelled data. In this paper, we address this issue by proposing Guided Adversarial Autoencoder (GAAE), which can generate superior conditional audio samples from unlabelled audio data using a small percentage of labelled data as guidance. Representation learned from unlabelled data without any supervision does not guarantee its' usability for any downstream task. On the other hand, during the representation learning, if the model is highly biased towards the downstream task, it losses its generalisation capability. This makes the learned representation hardly useful for any other tasks that are not related to that downstream task. The proposed GAAE model also address these issues. Using this superior conditional generation, GAAE can learn representation specific to the downstream task. Furthermore, GAAE learns another type of representation capturing the general attributes of the data, which is independent of the downstream task at hand. Experimental results involving the S09 and the NSynth dataset attest the superior performance of GAAE compared to the state-of-the-art alternatives.

\end{abstract}

\begin{keywords}
Audio Generation, Representation Learning, Generative Adversarial Neural Network, Guided Generative Adversarial Autoencoder
\end{keywords}

\titlepgskip=-15pt

\maketitle

\section{Introduction}
\label{sec:introduction}

\PARstart{R}{epresentation} learning aims to map higher-dimensional data into a lower-dimensional representation space where the variational factors of the data are disentangled. Learning a disentangled representation from an unlabelled dataset opens a window of opportunity for researchers to utilise the vastly available unlabelled data for any downstream tasks \cite{francesco_2019}. Such as, a representation learnt from freely available YouTube audios (movie, news etc.) can be used to improve 
%BS
a task such as emotion recognition 
from audio where a large labelled dataset is unavailable.

% Representation learning aims to map higher-dimensional data into a lower-dimensional representation space where the variational factors of the data are disentangled. Learning a disentangled representation from an unlabelled dataset opens a window of opportunity for researchers to utilise the vastly available unlabelled data for any downstream tasks \cite{francesco_2019}. Such as, a representation learnt from freely available YouTube audios (movie, news etc.) can be used to improve 
% %BS
% a task such as emotion recognition 
% from audio where a large labelled dataset is unavailable.

Generative Adversarial Neural Network (GAN) \cite{goodfellow:2014} has shown great promise for learning powerful representation. GAN is comprised of a Generator network and a Discriminator network, where these networks are trained to defeat each other based on a minimax game. During training, the Generator tries to fool the Discriminator by generating real-like samples from a random noise/latent distribution, and the Discriminator tries to defeat the Generator by differentiating the generated sample from the real samples \cite{goodfellow:2014}. During this game-play, the Generator disentangles the underlying attributes of the data in the given random latent distribution \cite{radford2015}. This 
helps in learning powerful representations \cite{chen2016infogan, chorowski_wavenet_autoencoder, radford2015, donahue2019large, zhao:2017, makhzani:2016, karras2019style} in a unsupervised manner. GAN based models pose great promise in audio research where limited or no labelled data is available.

The representation learning performance of the GANs usually improves along with its' sample generation quality. Intuitively, GAN models that can generate high-quality samples, intrinsically learns powerful representation \cite{donahue2019large}. GAN-based models are successful at generating high-fidelity images, however, they fail to perform likewise for the complex audio waveform generation as it requires modelling higher-order temporal scales \cite{engel2019gansynth}. To successfully generate audio with GANs, many researchers have worked with the spectrogram of the audio which can be converted back to the audio with minimal loss \cite{chris_wspecgan, engel2019gansynth, marafioti2019adversarial}.  Recently proposed high performing GAN architectures such as BigGAN \cite{Andrew_biggan} and StyleGAN \cite{karras2019style} are not well explored in the audio field, leaving a room to explore the compatibility of these models for audio data. 

A representation learnt with GANs in a completely unsupervised manner does not guarantee the usability of the learnt representation for any particular downstream task. This is because it can ignore the important characteristics of the data during the training which is important for succeeding in the downstream task \cite{haque2020guided}. So, some bias towards the downstream task is necessary during the unsupervised training to succeed in that downstream task \cite{francesco_2019}.

GAN models perform better for conditional generation using labelled data. The labels add useful side information during the training, which helps the GAN models to decompose overall sample generation tasks into sub-tasks according to the conditioned labels. Though the conditional generation helps to improve performance significantly, it requires an enormous amount of labelled data \cite{lucic2019highfidelity}, which is costly and/or error-prone. Using the GAN models to generate high-quality samples with a minimum amount of labelled data therefore remains a crucial challenge \cite{haque2020guided}.

% Here, learning a powerful representation using GAN models for a particular task, from an unlabelled dataset, require good conditional generation as well as some guidance towards the downstream task. 
%We addressed this 
In our previous work, we propose a BigGAN based architecture called ``Guided Generative Adversarial Neural Network (GGAN)'', which can generate state-of-the-art (SOTA) conditional audio with fewer labelled data. This labelled data is used as a guidance to force GGAN to learn guided representation for any downstream task at hand. Note that, the learned representation for any particular downstream task makes it less useful for any other task that is unrelated to the downstream task \cite{haque2020guided}. In many cases, it is desirable to learn representation in a manner so that it can be used for any particular downstream task as well as can be used for any future tasks independent of the downstream task at hand \cite{bengio:2013}.
It is a challenging problem to learn both generalised and guided representation at the same time with conditional GAN architectures. During the training of any conditional GAN, the latent noise/samples are independent of the given condition. So, GAN learns to map the general characteristics of the training data from the latent samples, which is independent of the condition. On the other hand, if the condition is imposed on the latent samples/noise like GGAN, the latent cannot learn general characteristics as it is biased towards the conditioned attributes. In this paper, we address this problem. 
% and the gap in the
% superior conditional audio generation research by proposing
% a novel autoencoder based model named “Guided Adversarial
% Autoencoder (GAAE)”. Here, our GAAE model can generate
% diverse and high-fidelity conditional audio samples with a minimal
% amount of labelled data as guidance. Moreover, using this superior generation quality; it can learn both general/style and guided representations from any unlabelled audio dataset.
Our contributions are as follows:

\begin{itemize}

%BS: PRESENT TENSE
\item We propose a novel autoencoder based GAN model GAAE, which can generate high-fidelity audio samples capturing the diverse modes of the training data distribution leveraging the guidance from a fewer labelled data samples from that dataset or a related dataset. 

\item We evaluate the conditional sample generation quality of the proposed model based on two audio datasets: the Speech Command dataset (S09) and the Musical Instrument Sound dataset (Nsyth). We demonstrate that the GAAE model performs significantly better than the SOTA models.

\item We achieve generalised and guided representation in our GAAE model. Evaluation results on three different datasets: the Speech Command dataset (S09), the Audio Book Speech dataset (Librispeech), and the Musical Instrument Sound dataset (Nsyth) show that the proposed GAAE model performs better than SOTA models.

\end{itemize}

\section{Background and Related Work}
\label{sec:background}

\subsection{Audio Representation Learning}
While there is a rich literature of supervised representation learning, due to our focus on unsupervised representation learning we will only discuss the related literature here. In the field of unsupervised representation learning, the self-supervised learning has become very popular recently due to its unprecedented success in the field of computer vision  \cite{zhang_colorful:2016, larsson:2016, doersch_un:2015, haque2018image, liu2019selfsupervised, Zhan_2019, feng2019self} and natural language processing  \cite{devlin2018bert,wu2019self,su2019vl, wang2019self}. Self-supervised learning uses information presents in the unlabelled data to create an alternative supervised signal to train the model for learning feature/representation. For an example, learning representation through predicting the rotation angel of images where rotation angel serves as supervised signal and this learned representation can be used to improve other related image classification tasks \cite{spyros:2018}. 

Likewise, in the audio field, researchers have achieved good performances using self-supervised representation learning. In their work, DeepMind  \cite{van:2018} have proposed a model to learn a useful representation from unsupervised speech data through predicting a future observation in the latent space. In another work from Google  \cite{de2019learning}, the representation is learnt by predicting the instantaneous frequency based on the magnitude of the Fourier transform.
Furthermore, Arsha et al.\ (2020)  \cite{A9054057} proposed a cross-modal self-supervised learning method to learn speech representation from the co-relationship between the face and the audio in the video. Other efforts have been made by researchers to learn a general representation by predicting the contextual frames of any particular audio frame like wav2vec  \cite{schneider2019wav2vec}, speech2vec  \cite{chung2018speech2vec}, and audio word2vec \cite{chung2016audio}. Likewise, there are other successful implementations  \cite{kawakami2020learning, riviere2020unsupervised, baevski2019vqwav2vec, baevski2019effectiveness} of the self-supervised representation learning in the field of audio. 

Though self-supervised learning is good for learning representations from unlabelled datasets, it requires manual endeavour to design the supervision signal  \cite{latif2020deep}. To avoid this, researchers have focused on fully unsupervised representation learning mainly using autoencoders \cite{amiriparian2017sequence,lee2009unsupervised, Xu_IEE}. In \cite{Neumann_2019}, the authors learnt representations with an autoencoder from a large unlabelled dataset, which improved the emotion recognition from speech audio. Similarly, in another work, the authors used a denoising autoencoder to improve affect recognition from speech data  \cite{ghosh2015learning}. Several works  \cite{W8268911, Chorowski_wavenet, hsu2019disentangling} have utilised Variational Autoencoders (VAEs)  \cite{kingma:2013} to learn an efficient speech representation from an unlabelled dataset. Recently, given the popularity of adversarial training, different works have been conducted by researchers to learn a robust representation with GANs  \cite{J7952656,yu2017adversarial} and Adversarial Autoencoders \cite{sahu2018adversarial, E7966273}. 

Though learning a representation from prodigiously available unlabelled datasets is very intriguing, the recent work from Google AI has proved that completely unsupervised representation learning is not possible without any form of supervision \cite{francesco_2019}. Also, representation learnt from an unsupervised method does not guarantee the usability of this learnt representation for any post use case scenario. Thus, 
%BS: 
as outlined, 
we proposed the Guided Generative Adversarial Neural Network (GGAN)  \cite{haque2020guided}, which can learn a powerful representation from an unlabelled audio dataset according to the supervision given from a fewer amount of labelled data. Therefore, in the learnt representation space, the GGAN disentangles attributes of the data according to the given categories from the labelled dataset, which benefits the related post-use case scenario. 

% Still, the generalisation is lost, and thus cannot be used for non-related tasks. For an example, if the GGAN is guided with a small amount of data with emotion labels and trained on a large number of speech audios from different people, the GGAN will learn an emotion-related representation ignoring the other attributes such as gender of the speaker, background noise, pitch, intensity etc. Therefore, this will help to improve the emotion recognition task but cannot be 
% %BS: 
% fully 
% used for other tasks such as speaker gender identification  \cite{haque2020guided}. Hence, we overcome this shortcoming by proposing
% %BS: 
% ---as outlined---
% the Guided Adversarial Autoencoder (GAAE) model, which can learn general attributes of an unlabelled dataset in the representation space as well as the characteristics according to the given guidance from the fewer labelled data samples.   

\subsection{Audio Generation}

Most of the audios are periodic, and high-fidelity audio generation requires modelling a higher order magnitude of the temporal scales, which makes it a challenging problem  \cite{engel2019gansynth}. Most of the research works related to audio generation are based on the audio synthesis viz; Aaron and et al.\ (2016) have proposed a powerful autoregressive model named ``Wavenet'', which works very well on text to speech (TTS) synthesis for both English and Mandarin. Later, the authors have improved this work by proposing ``Parallel Wavenet'', which is 20 times faster than the original Wavenet. Other researchers have utilised the seq2seq model for TTS such as Char2Wav \cite{sotelo2017char2wav} and TACOTRON  \cite{wang2017tacotron}. However, these audio generation methods are conditioned on the text data and mainly focused on speech generation. Thus, these methods cannot be generalised to all other audio domains, even for speech data where transcripts are not available.

In the context of generating audio without any condition on the text data, the GANs are very promising due to their massive success in the field of computer vision  \cite{donahue2019large, dumoulin2016adversarially, DonahueKD16, karras2019analyzing, karras2019style}. However, porting these GAN architectures directly to the audio domain does not offer similar performance as the audio waveform is mostly more complex than an image  \cite{chris_wspecgan,engel2019gansynth}. Therefore, researchers have focused on generating spectrogram (2D image-like representation of audio) rather than generating directly a  waveform. Then, the generated spectrogram is converted back to audio. Chris et al.\ (2019)  \cite{chris_wspecgan} have trained a GAN-based model to generate spectrograms and successfully converted them back to the audio domain with the Griffin-Lim algorithm  \cite{griffin1984signal}.
In their TiFGAN paper  \cite{marafioti2019adversarial}, the authors have proposed a phase-gradient heap integration (PGHI)  \cite{zden_25} algorithm for better reconstruction of the audio from the spectrogram with minimal loss. As the PGHI algorithm is good at reconstructing audio from the spectrogram, now the challenge is to generate a realistic spectrogram. As the spectrogram is---as outlined---an image-like representation of the audio,  any GAN based framework from the image domain should be compatible.  Hence, the BigGAN architecture  \cite{Andrew_biggan} has shown promising performance at generating conditional high resolution/fidelity images, but it was not well explored for audio generation. 
%BS: This is the THIRD TIME you say this - I would delete it, as it becomes tiring :) I did rephrase a bit:
In this paper we address this gap.

\subsection{Closely Related Architectures}

The proposed GAAE model is a semi-supervised model, as we leverage a small amount of labelled data during the training. In \cite{Spurr2017}, the authors proposed a semi-supervised version of the InfoGAN model  \cite{chen2016infogan} to capture a specific representation and generation according to the supervision which comes from a small number of labelled data. But, the success of this model in terms of the complex data distribution is not evident. Other researchers have explored the scope of semi-supervision in GAN architectures  \cite{springenberg2015unsupervised, sricharan2017semisupervised, lucic2019highfidelity} to improve the conditional generation, but most of these works are not explored in the audio domain which leaves a major gap for the researchers to address. The GAAE model is based on an Adversarial Autoencoder (AAE)  \cite{makhzani:2016}, where we have extended the AAE model to learn both guided and generalise/style representation from an unlabelled dataset in a semi-supervised fashion. Furthermore, in the GAAE model, we have implemented a unique way to leverage the small amount of labelled data for conditional audio generation. Here, we have also proposed a way to utilise the generated conditional samples for improving the representation learning during the training. Moreover, the building block for our GAAE model is a BigGAN architecture; thus, we further contribute by exploring the use of a BigGAN in an autoencoder-based model for audio data.

 %BS finished editing

\section{Proposed Research Method}
\label{sec:proposed_method}

% \begin{figure*}[ht]
%     \centering
%     \frame{\includegraphics[width=.6 \linewidth]{GAE_model.jpg}}
%     \caption{This figure illustrates the overall architecture of the GAAE model. Different networks of the GAAE model are shown along with the connections between them. In the figure, the arrows are coloured to highlight the flow of any input/output of the model. For the discriminator, the red boxes show the fake samples and the green boxes indicate  the real samples. Here, $x_{u}$ is the unlabelled data sample, $x_{l}$ is the labelled data sample,$\hat{x}_{u}$ is the reconstructed data sample, $y_{r}$ is the random conditions, and $z$ is the known latent distribution.}
%     \label{fig:model_architecture}
% \end{figure*}

\Figure[!t]()[width=0.90\textwidth]{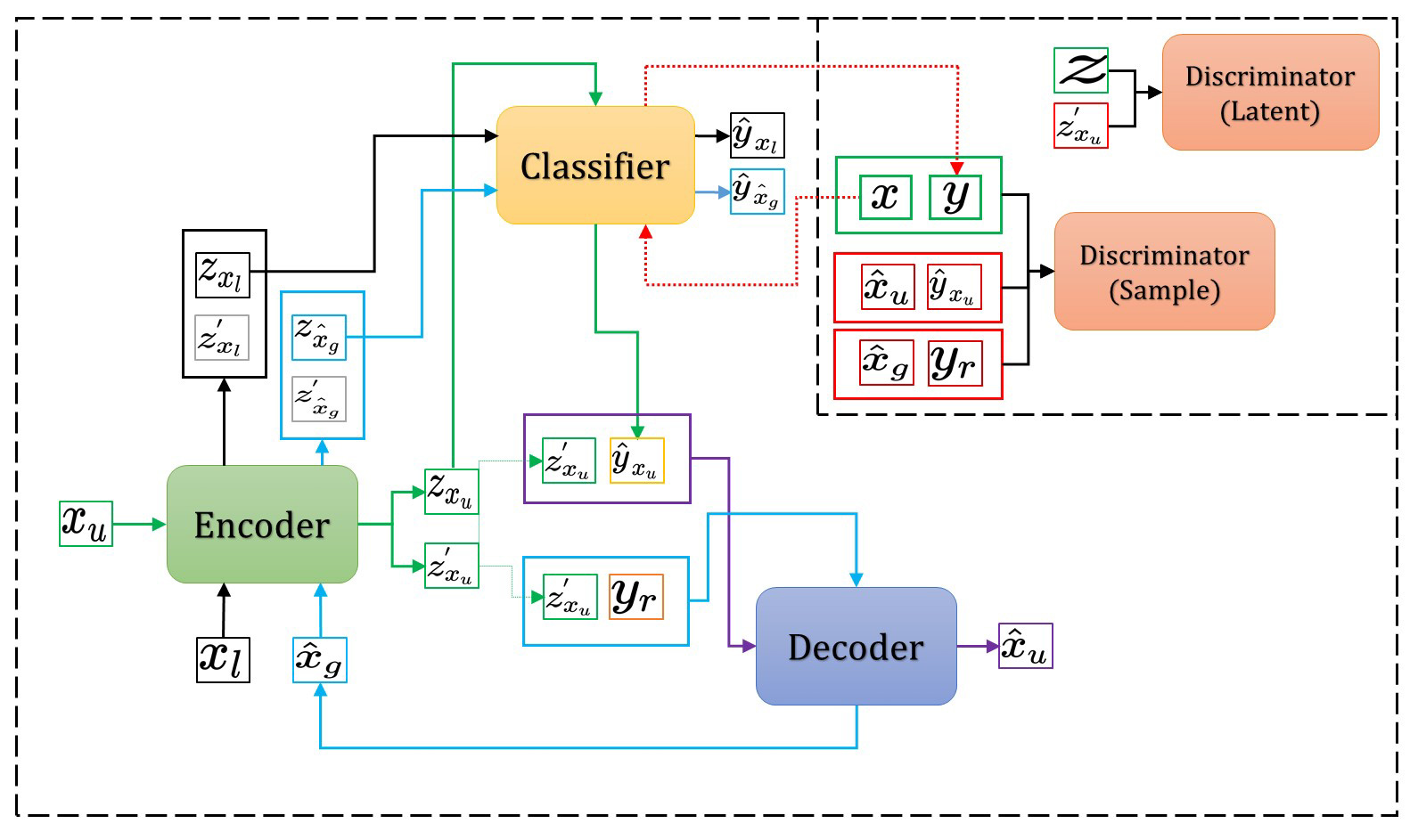}
   {This figure illustrates the overall architecture of the GAAE model. Different networks of the GAAE model are shown along with the connections between them. In the figure, the arrows are coloured to highlight the flow of any input/output of the model. For the discriminator, the red boxes show the fake samples and the green boxes indicate  the real samples. Here, $x_{u}$ is the unlabelled data sample, $x_{l}$ is the labelled data sample,$\hat{x}_{u}$ is the reconstructed data sample, $y_{r}$ is the random conditions, and $z$ is the known latent distribution.\label{fig:model_architecture}}

\subsection{Architecture of the GAAE}

The GAAE consists of five neural networks: the Encoder $E$, the Decoder $D$, the Classifier $C$, the Latent Discriminator $L$ and the Sample Discriminator $S$. Let the parameters for these networks be $\theta_{e}$, $\theta_{d}$, $\theta_{c}$, $\theta_{L}$, and $\theta_{S}$ respectively. Figure \ref{fig:model_architecture} shows the whole architecture of the model and the description is as follows. 

\subsubsection{Encoder}

The Encoder $E$ takes any unlabelled data sample $x_{u}$ $\sim$ $p_{data}$ and outputs two latent samples $z_{x_{u}}$ $\sim$ $u_z$ and $z^{'}_{x_{u}}$ $\sim$ $q_z$, where $p_{data}$ is the true unlabelled data distribution, and $u_z$,$q_z$ are two different continuous distributions learned by the $E$. We require the latent $z_{x_{u}}$ to capture the guided attributes/characteristics  of the data and the latent $z^{'}_{x_{u}}$ to capture the general/style attributes of the data.

\subsubsection{Classifier}

We have a classifier network $C$ which is trained with limited labelled data  $x_{l}$ $\sim$ $p_{ldata}$, where $p_{ldata}$ is the labelled data distribution and not necessarily $p_{ldata}$ $\subset$ $p_{data}$. Here, with this $p_{ldata}$, the whole model gets guidance---thus, we call this data as ``guidance data''. Now, the $C$ network takes any latent sample and predicts the category class for that latent sample. To train $C$, we pass $x_{l}$ through the $E$ network and get two latent vectors \{$z_{x_{l}}$,$z^{'}_{x_{l}}$\} = $E(x_{l};\theta_{e})$. Then, we only forward $z_{x_{l}}$ through $C$ to get the predicted label $\hat{y}_{x_{l}}$ = $C(z_{x_{l}};\theta_{c})$ and train $C$ against the true label $y_{l}$ $\sim$ $Cat(y_{l},k = n)$ of the sample $x_{l}$, where $Cat(y_{l},k = n)$ is the categorical distribution with $n$ numbers of categories/labels. These labels are used as one-hot vector. For now, lets consider that $C$ can classify the label of any sample correctly.

\subsubsection{Decoder}
The Decoder $D$ maps any latent and categorical class/label variable to the data sample. Now, to get the reconstructed sample of $x_{u}$, we pass the latent $z^{'}_{x_{u}}$ and the label of $x_{u}$ through the $D$ network. As $x_{u}$ is an unlabelled data sample, we get the label $\hat{y}_{x_{u}}$ = $C(z_{x_{u}},\theta_{c})$ through the network $C$ and obtain the reconstructed sample $\hat{x}_{u}$ = $D(z^{'}_{x_{u}},\hat{y}_{x_{u}} ; \theta_{d})$ from the $D$ network. 
Here, we also want to use the $D$ network for generating samples according to the given condition along with the reconstruction. Therefore, the same latent $z^{'}_{x_{u}}$ is used with a random categorical variable (one-hot vector) $y_{r}$, sampled from categorical distribution $Cat(y_{r}, K = n,p = \frac{1}{n})$ , where $n$ is the number of categories/labels, and the sampling probability for each category is $\frac{1}{n}$. Now, we obtain the generated sample $\hat{x}_{g}$ $\sim$ $p_{gdata}$, where $p_{gdata}$ is the generated data distribution by the $D$ network, and it is trained to match $p_{gdata}$ with the true data distribution $p_{data}$. Here, the size of $n$ is the same as of the guided data, and we want the $D$ network to generate data according to the categories from the guided data. Therefore, we ensure this with the Discriminator where the Discriminator receives the labels of the data from the network $C$.  
As we use a small number of labelled data, it is hard to train $C$ due to the problem of overfitting. Hence, we use the generated sample $\hat{x}_{g}$ and train the $C$ network considering $y_{r}$ as the true label/category, where the predicted label is $\hat{y}_{\hat{x}_{g}}$ = $C(E(\hat{x}_{g},\theta_{e}),\theta_{c})$. 

Here, $C$ depends on the correct conditional generation from $D$, and $D$ depends on the classification from the network $C$. During the training, the $C$ network starts to predict the category of some samples from the given labelled data correctly. Likewise, the Discriminator learns to identify the correct category for those samples and forces the $D$ network to generate samples with the attributes related to these correctly classified samples. These generated samples bring more characteristics with them, which are not present in the given labelled data but belong to the data distribution. Now, as we feed these generated samples again to the $C$ network with the associated conditional categories as correct labels, it learns to predict the correct category for more samples related to that generated samples. Then again, these new correctly classified samples improve the conditional generation of the $D$ network. Hence, throughout the training, the $C$ network and $D$ network improve each other continuously. Meanwhile, during the training, the representation learning (latent generation) capability of the $E$ network is also ameliorated via the process of reconstructing sample $x_{u}$, which also improves the performance of the $C$ and $D$ network eventually.   

\subsubsection{Discriminators}

The GAAE model has two discriminators: the Sample Discriminator $S$ and the Latent Discriminator $L$. $S$ makes sure that the generated sample $\hat{x_{g}}$ and the reconstructed sample $\hat{x_{u}}$ match the sample from the true data distribution $p_{data}$. We train $S$ with the sample and its label. Now, for the samples $\hat{x_{g}}$ and $\hat{x_{u}}$, we have the labels $y_{r}$,$\hat{y}_{x_{u}}$ respectively. Hence, the pairs  $(\hat{x_{g}},y_{r})$ and $(\hat{x_{u}},\hat{y}_{x_{u}})$ are considered fake labels for the discriminator $S$. For the true data, both $x_{l}$ and $x_{u}$ are used together, where we get the label for the sample $x_{u}$ from $C$, and, for the sample $x_{l}$ we use the available true labels. Hence, in terms of distribution perspective, we obtain the data distribution $p_{mdata}$, mixing the distributions $p_{ldata}$ and $p_{data}$. Accordingly, $S$ is trained with the true sample data $x$ $\sim$ $p_{mdata}$ along with its associated label $y$ if it  exists, otherwise with the predicted label from $C$.

Here, the network $E$ learns to map the general characteristics of the data onto the latent distribution $q_z$, excluding the categories from the guided data. Now, if we can draw the sample from the  $q_z$ distribution, then, by using the categorical distribution as condition, we can generate diverse data for different categories (categories from the guided data) from the Decoder $D$. We can only sample from $q_z$, if the distribution is known to us. Therefore, we use another Discriminator $L$ so that the $E$ network is forced to match $q_z$ to any known distribution $p_z$, where $p_z$ can be any known continuous random distribution (e.\,g.,  Continuous Normal Distribution, or Continuous Uniform Distribution). The $L$ network is trained through differentiating between the true latent $z$ $\sim$ $p_z$ and the fake latent $z{'}_{x_{u}}$.  

\subsection{Losses}

\subsubsection{Encoder, Classifier and Decoder}

For the $E$ and $D$ networks, we have the sample generation loss $G_{loss}$, the sample reconstruction loss $R_{loss}$, and the latent generation loss $L_{loss}$.
To calculate the generation and discrimination loss, we use hinge loss, and for the reconstruction loss the Mean Squared Error (MSE) loss. For the $G_{loss}$, we take the average of the generation loss for $\hat{x_{u}}$ and $\hat{x_{g}}$. Therefore,   

\begin{equation}
\label{eq:1}
\begin{aligned}
G_{loss} = -\frac{1}{2}(S(\hat{x_{u}},\hat{y}_{x_{u}} ; \theta_{s}) + S(\hat{x_{g}},y_{r}); \theta_{s}).\\
\end{aligned}
\end{equation}

\begin{equation}
\label{eq:2}
\begin{aligned}
L_{loss} = -(L( z^{'}_{x_{u}}; \theta_{l}).\\
\end{aligned}
\end{equation}

\begin{equation}
\label{eq:30}
\begin{aligned}
R_{loss} = \frac{1}{N}\sum\limits_{i=1}^N{ (\hat{x}_{u_{i}} - x_{u_{i}})^{2}}.\\
\end{aligned}
\end{equation}

Now, for the $C$ network, we calculate the classification loss $Cl_{loss}$, $Cg_{loss}$ for the labelled data sample $x_{l}$ and the generated sample $\hat{x_{g}}$ respectively. Here, $\hat{x_{g}}$ is used as a constant, so it is considered like a sample data $x_{l}$. We only forward propagate $x_{u}$ through $E$ and $D$ and no gradient is calculated for generating $\hat{x_{g}}$ when it is only used for the loss $Cg_{loss}$. The model is implemented with pytorch \cite{paszke2019pytorch} and we detach the gradient of $x_{g}$ when $Cg_{loss}$ is calculated. Therefore,

\begin{equation}
\label{eq:3}
\begin{aligned}
Cl_{loss} = - \sum y_{l} \log \hat{y}_{x_{l}}.\\
\end{aligned}
\end{equation}

\begin{equation}
\label{eq:4}
\begin{aligned}
Cg_{loss} = - \sum y_{r} \log \hat{y}_{\hat{x}_{g}}.\\
\end{aligned}
\end{equation}

We get the a combined loss $EDC_{loss}$ for $E$,$D$ and $C$. The $EDC_{loss}$ is calculated as 

\begin{equation}
\label{eq:5}
\begin{aligned}
EDC_{loss} = \alpha \cdot (\omega_{1} \cdot G_{loss} + \omega_{2} \cdot ( \lambda \cdot R_{loss})) + 
\\ \beta \cdot (\omega_{3} \cdot Cl_{loss} + \omega_{4} \cdot Cg_{loss} + \omega_{5} \cdot L_{loss}).\\
\end{aligned}
\end{equation}

Here, the weights of the $E$,$C$, and $D$ networks are updated to minimise the loss $EDC_{loss}$, where $\omega_{1}$, $\omega_{2}$, $\omega_{3}$, $\omega_{4}$, $\omega_{5}$, $\alpha$, $\beta$, and $\lambda$ are the hyperparameters. The successful training of our GAEE model depends on these parameters. At the beginning of the training, we noticed that the value of $R_{loss}$ falls rapidly compared to other losses and results in a very small gradient value. To mitigate this problem, we multiply $R_{loss}$ with a hyperparameter $\lambda$ $\in$ $\mathbb{R}_{> 0}$ and after hyperparameter tuning, we found 20 as an optimal value for  $\lambda$. The $D$ network of the model is tuned for both the reconstruction loss $R_loss$ and the generation loss $G_loss$. Therefore, to balance between these two losses, the hyperparameter $\omega_{1}$ and $\omega_{2}$ is used where $\omega_{1}$, $\omega_{2}$ $\in$ $[0,1]$ and $\omega_{1}$ + $\omega_{2}$ = 1. Here, we can force the model to focus more on either loss by increasing the hyperparameter for that particular loss. Likewise, for $Cl_{loss}$, $Cg_{loss}$ and $L_{loss}$, we use the hyperparameters $\omega_{3}$, $\omega_{4}$, $\omega_{5}$ respectively, where  
$\omega_{3}$, $\omega_{4}$, $\omega_{5}$ $\in$ $[0,1]$ and $\omega_{3}$ + $\omega_{4}$ + $\omega_{5}$ = 1. In the $EDC_{loss}$, $G_{loss}$ and $R_{loss}$ are responsible for the sample generation quality, where  $Cl_{loss}$, $Cg_{loss}$ and $L_{loss}$ are responsible for the latent generation quality. So, to balance between sample generation and latent generation, we use two hyperparameters $\alpha$ and $\beta$, where $\alpha$, $\beta$ $\in$ $[0,1]$, and  $\alpha$ + $\beta$ = 1.

\subsubsection{Discriminators' loss}

For the Discriminators $S$ and $L$, we use hinge loss. The discrimination loss for the fake samples are averaged as we calculate the loss for both $\hat{x}_{u}$ and $\hat{x}_{g}$. Let the discrimination loss for $S$ and $L$ be $S_{loss}$, $L_{loss}$ respectively. Therefore, 

\begin{equation}
\label{eq:6}
\begin{aligned}
S_{loss} = - min(0, -1 + S(x, C(E(x,\theta_{e});\theta_{c}) ; \theta_{s})) \\
-  \frac{1}{2} (min(0, -1 - S(\hat{x}_{u}, \hat{y}_{x_{u}}; \theta_{s})) \\+ min(0, -1 - S(\hat{x}_{g}, \hat{y}_{r}; \theta_{s}))). \\
\end{aligned}
\end{equation}

\begin{equation}
\label{eq:7}
\begin{aligned}
L_{loss} = - min(0, -1+ L(z,\theta_{l}) \\- min(0,-1-L(\hat{z}_{x_{u}},\theta_{l})). \\
\end{aligned}
\end{equation}

Here, we update the parameters $\theta_{s}$ and $\theta_{l}$ to maximise the loss  $S_{loss}$ and $L_{loss}$ respectively.  Algorithm 1 shows the training mechanism for the GAAE model. 

\begin{algorithm}[t!]
\caption{\small Minibatch stochastic gradient descent training of the proposed GAAE model. The discriminator is updated $k$ times in one iteration. Here, for our experiment, we use  $k=2$ for better convergence.}

\begin{algorithmic}[1]
\label{alg:AGF1}
\FOR{number of training iterations}
  \FOR{$k$ steps}
    
    \STATE { Sample the latent/noise samples $ \{ \bm{z^{(1)}}\dots, \bm{z^{(m)}} \} $ from $p_z$, the conditions (labels) $ \{ \bm{y_{r}^{(1)}}, \dots, \bm{y_{r}^{(m)}} \} $ from $Cat(y_{r})$, the unlabelled data samples  $ \{ \bm{x_{u}^{(1)}}, \dots, \bm{x_{u}^{(m)}} \}$ from $p_{data}$ and the labelled data samples  $ \{ \bm{x_{l}^{(1)}}, \dots, \bm{x_{l}^{(m)}} \}$ from $p_{ldata}$. Here, $m$ is the minibatch size.}
    
    \STATE {Update the discriminator $S$ by ascending its stochastic gradient:
        \[
            \nabla_{\theta_{s}} \frac{1}{m} \sum_{i=1}^m \left[\bm{{S_{loss}}}^{(i)}\right].
        \]}   
        
    \STATE {Update the discriminator $L$ by ascending its stochastic gradient:
        \[
            \nabla_{\theta_{l}} \frac{1}{m} \sum_{i=1}^m \left[\bm{{L_{loss}}}^{(i)}\right].
        \]}   
  \ENDFOR
   
    \STATE {Repeat step [3].}
    
    \STATE {Update the Encoder $E$, Decoder $D$, and Classifier $C$ by descending its stochastic gradient:
        \[
            \nabla_{\theta_{e},\theta_{d},\theta_{c}} \frac{1}{m} \sum_{i=1}^m \left[\bm{{EDC_{loss}}}^{(i)} \right].
        \]}
     \ENDFOR

% \algstore{myalg}
\end{algorithmic}
\end{algorithm}

 %BS finished editing

\section{Data and Evaluation Metrics}
\subsection{Datasets}

The effectiveness of the GAAE model is evaluated on both speech and non-speech audios. For the speech audio, we chose the S09 dataset \cite{Pete_03209} and the Librispeech dataset \cite{panayotov2015librispeech}. For the non-speech audio, we use the popular Nsynth dataset 
\cite{nsynth2017}. The S09 dataset consists of utterances for different digit categories from zero to nine. This dataset comprises 23,000 one-second audio samples uttered by 2618 speakers, where it only contains the labels for the audio digits \cite{Pete_03209}.

The Librispeech dataset is an English speech dataset with 1000 hours of audio recordings, and there are three subsets available in the Librispeech dataset containing approximately 100, 300, and 500 hours of recordings, respectively. For our work, we use the subset with 100 hours of clean recordings. In this subset, the audios are uttered by 251 speakers where 125 are female, and 126 are male \cite{panayotov2015librispeech}. For our experiment, we only apply the audios along with the gender labels of the speakers. 

The Nsynth audio dataset contains 305,979 musical notes of size four seconds from ten different instruments,  where the sources are either acoustic, electronic, or synthetic \cite{nsynth2017}. We use three acoustic sources: Guitar, Strings, and Mallet from the Nsynth to test the compatibility of the GAAE model for a non-speech dataset.

\subsection{Data Preprocessing}

We use the audio of length one second and the sampling rate of 16kHz. For the Librispeech dataset, the one-second audio is taken randomly from any particular audio clip where for the Nsynth dataset, the first one-second is taken from any audio sample as it holds the majority of the instrument sound representation.

The audio data is converted to the log-magnitude spectrograms with the short-time Fourier Transform, and the generated log-magnitude spectrograms of the GAAE model are converted to audio using the PGHI algorithm \cite{zden_25}. In the rest of the paper, we refer to the log-magnitude spectrogram as the spectrogram.

To obtain the spectrogram representation of the audio we followed the procedure from this paper \cite{andrTIFGAN}. The short-time Fourier Transform is calculated with an overlapping Hamming window of size 512\,ms, and the hopping length 128\,ms. Therefore, we get the size of the spectrogram as 256 $\times$ 128, 1D matrix. We standardise the spectrogram with the equation $\frac{X-\mu}{\sigma}$, where $X$ is the spectrogram, $\mu$ is the mean of the spectrogram, and $\sigma$ is the standard deviation of the spectrogram. We clip the dynamic range of the spectrogram at $-r$, where, for the S09 and Librispeech dataset, we determine the suitable value of $r$ to be $10$, and for the Nsynth dataset we determine it $15$. Here, the log-magnitude spectrograms is a normal distribution and any inappropriate value of the $r$ can make the distribution skewed, which is not appropriate for training the GAAE network. We investigate the histogram of the values combining all the log-magnitude spectrograms from the whole training dataset to determine the value of $r$. After the clipping, we normalise the spectrogram values between $-1$ and $1$. The spectrogram representation of the audio is used as the input to the GAAE model, which then generates spectrograms with values between -1 and 1. We then convert these spectrograms to audios via the PGHI algorithm. In this paper we refer to these audios calculated from generated spectrograms as ``generated audios''.

\subsection{Measurement Metrics}
We measure the performance of the GAAE model based on the generated samples and the learnt representations. The generated samples are evaluated with the Inception Score (IS) \cite{salimans:2016} and Fr\'echet Inception Distance (FID) \cite{heusel2017gans,barratt2018note}, which have become a de-facto standard for measuring the performance of any GAN based model \cite{shmelkov}. 

To evaluate the representation/latent learning, we consider classification accuracy, latent space visualisation,  and latent interpolation.

\subsubsection{Inception Score (IS)}

The IS score is calculated based on the pretrained Inception Network \cite{szegedy2014going} trained on the ImageNet dataset \cite{imagenet_cvpr09}. The logits are calculated for the images from the bottleneck layer of the Inception Network. Then, the score is calculated using
\begin{equation}
\label{eq:iscore}
\begin{aligned}
\exp(\mathbb{E}_{x} {KL}(p(y|x) || p(y))). \\
\end{aligned}
\end{equation}
Here, $x$ is the image sample, $KL$ is the Kullback-Leibler Divergence (KL-divergence) \cite{kl}, $p(y|x)$ is the conditional class distribution for sample $x$ predicted by the Inception Network, and $p(y)$ is the marginal class distribution. The IS score computes the KL-divergence between the conditional label distribution and the marginal label distribution,  {\bf where the higher value indicates good generation quality.}

\subsubsection{Fr\'echet Inception Distance (FID)}
The IS score is computed solely on the generated samples; thus, no comparison is made between the generated and real samples which is not a good measure for the samples' diversity (mode) of the generated samples. The FID score solves this problem by comparing real samples with the generated samples \cite{shmelkov} during the score calculation. The Fr\'echet Inception Distance (FID) computes the Fr\'echet Distance \cite{dowson1982frechet} between two multivariate Gaussian distributions for the generated and real samples, parameterised by the mean and the covariance of the features extracted from the intermediate layer of the pretrained Inception Network. The FID score is calculated using 
\begin{equation}
\label{eq:fidscore}
\begin{aligned}
||\mu_r - \mu_g||^2 + \text{Tr} (\Sigma_r + \Sigma_g - 2 (\Sigma_r \Sigma_g)^{1/2}), \\
\end{aligned}
\end{equation}
where, $\mu_{r}$, $ \mu_{g}$  are the means for the features of the real and generated samples,  respectively, and similarly, $\Sigma_{r}$, $\Sigma_{g}$ are the covariances, respectively. {\bf A lower value of the FID score indicates good generation quality.} 

The Inception Network is trained on the imagenet dataset, thus, offering reliable IS and FID scores for a related image dataset, but the spectrograms of the audios are entirely different from the imagenet samples. So, the Inception Network does not offer trustworthy scores for the audio spectrograms. Hence, instead of using the Inception model, we train a classifier network based on the audio datasets and use this trained classifier to calculate the IS and FID scores. For S09 dataset, we use the pretrained classifier released by the authors of the paper ``Adversarial Audio Synthesis'' \cite{chris_wspecgan}. For the Nsynth dataset, we train a simple Convolutional Neural Network (CNN) as the Classifier, as there was no pre-trained classifier available.

\section{Experimental Setup, Results and Discussion}
For implementing our GGAN model, we follow the network implementations, optimisation, and hyperparameters from the BigGAN paper \cite{Andrew_biggan}. For the optimisation, we use the Adam optimiser \cite{Kingma2015AdamAM}.  Learning rate of $5 \cdot 10^{-5}$ is used for the networks $E$, $D$, and $C$, where $2 \cdot 10^{-4}$ is the learning rate for both $S$ and $L$. Details of the network architectures are given in the appendix A (Architectural Details).

\subsection{Impact of Labelled Data for Conditional Sample Generation}

\subsubsection{Setup}
First, we evaluate the conditional sample generation quality (measured with IS and FID score) of the GAAE model for different percentage of labelled data (1\% - 5 \%, 100\%) as guidance. 

The IS and FID scores is calculated based on the 50,000 generated samples \cite{salimans:2016} for the random latent $z$, and the random condition $y_{r}$. The spectrograms of the samples are generated using the Decoder $D$ network and converted to audios. These generated audios are then used to calculate the IS and FID scores. For all the datasets, we use a continuous normal distribution of size 128 to sample the latent $z \sim \mathcal{N}(\mu = 0,\,\sigma^{2} = 1) $. For the S09 dataset, we use the ten digit categories (0-9) as the conditions $y_{r} \sim Cat(y_{r}, K = 10,p = 0.1)$. We use the three instrument categories (1-3) as conditions $y_{r} \sim Cat(y_{r}, K = 3,p = 0.33)$ for the Nsynth dataset. 

For any percentage of data used as guidance, we train the GAAE model three times. Each training takes approximately 60,000 iterations with mixed-precision \cite{micikevicius2017mixed} for the batch size 128.  Each time, a dataset is sampled randomly for guidance. Rest of the data is used as unsupervised manner. We limited ourselves to three times due to having high wall time: approximately 21 hours on the two Nvidia p100 GPUs. The total wall time for the S09 and the Nsynth dataset is approximately 21 $\times$ 3 $\times$ 6 (1-5\%,100\% data) $\times$ 2 (two datasets) = 756 hours or 31.5 days.

The results of the GAAE model are compared with a Supervised BigGAN \cite{brock_20118_bigGan} and an Unsupervised BigGAN \cite{brock_20118_bigGan}. For the S09 dataset,  we take the results from the GGAN publication \cite{haque2020guided}.  For the Nsynth dataset, we train these models with a similar setting as was used in the GGAN paper. To calculate the IS and FID score for the Nsynth dataset, we use our pretrained supervised CNN classifier (details in the appendix A) trained on three classes: Guitar, Strings, and Mallet.

\subsubsection{Results and Discussions}
The percentage of labelled training data used as guidance has a significant impact on the IS and FID score, which can be found from the table \ref{tab:table5}. The more we feed the labelled data during the training, the more we boost the performance of the GAAE model for sample generation and diversity. However, notably only with 1\% labelled data, the GAAE model achieves acceptable performance. For 5\,\% labelled data, GAAE achieves scores close to that of using 100\% labelled data. So, we compare the scores for 5\% data, with other models in the literature.

The results for S09 dataset are summarised in Table~\ref{tab:table1}. Using only 5\,\% labelled training data as guidance, the GAAE model achieves  IS score $7.28 \pm 0.01$ and FID score of $22.60 \pm 0.07$. The IS score of GAAE is close to that produced by the supervised BigGAN model ($7.33 \pm 0.01$) and better than other models mentioned in table \ref{tab:table1}. 
Even the GAAE model has outperformed the supervised BigGAN model (FID score: $24.40  \pm 0.50$) in terms of diverse image generation, where the GAAE has used only 5\,\% labelled data and the supervised BigGAN is trained with all available labelled training data.

For the Nsynth dataset, the GAAE model has achieved the IS score of $2.58 \pm 0.03$ and the FID score of $141.71 \pm 0.32$ again with 5\,\% labelled training data as guidance. Performance of GGAN in terms of IS score is very close to that of the supervised BigGAN ($2.64 \pm 0.08$) and better than that of the unsupervised BigGAN ($2.21 \pm 0.11$). The performance in terms of FID score is even better than that of the supervised BigGAN ($148.30 \pm 0.23$). Table \ref{tab:table2} presents the comparisons. 

The decoder is trained for both reconstruction and generation of the training data. During the reconstruction, it tries to reconstruct all the training samples, which helps it to learn more modes of the data distribution than the supervised BigGAN model. Figure~\ref{fig:nsyth_all_samples} and \ref{fig:real_generated_samples} display the spectrogram of the generated and the real samples of the Nsynth, S09 datasets, respectively.  From these figures, we observe that the generated samples are visually indistinguishable from the real samples. This attests the superior generation quality of the GAAE model. This is also true when we convert these spectrograms to audios. The audios can be found at: \href{https://bit.ly/3coz5qO}{{https://bit.ly/3coz5qO}}.

%%%%%%%%%%%%%%%%%%%%%%%%%%%%TABLE%%%%%%%%%%%%%%%%%%%%%%%%%%%%%%%%%%%

\begin{table}[t!]
\centering
% \caption{Comparison between the performance of the GAAE model and the other models on the S09 dataset, in terms of the quality of the generated samples, measured with IS and FID score.}

\caption{Comparison between the sample generation quality of the GAAE model and the other models for the S09 dataset. The generation quality is measured by IS score and FID scores.}
\label{tab:table1}
\begin{tabular}{|l|l|l|}\hline
\textbf{Model Name} & \textbf{IS Score} & \textbf{FID Score} \\ \hline
Real (Train Data) \cite{chris_wspecgan} & 9.18 $\pm$ 0.04 & - \\ \hline
Real (Test Data) \cite{chris_wspecgan} & 8.01 $\pm$ 0.24 &  -\\ \hline
TiFGAN \cite{andrTIFGAN} & 5.97 & 26.7 \\ \hline
WaveGAN \cite{chris_wspecgan} & 4.67 $\pm$ 0.01 &  -\\ \hline
SpecGAN \cite{chris_wspecgan} & 6.03 $\pm$ 0.04 &  -\\ \hline
{\bf Supervised BigGAN} & {\bf 7.33 $\pm$ 0.01} & {\bf 24.40  $\pm$ 0.50}  \\ \hline
Unsupervised BigGAN & 6.17 $\pm$ 0.20 & 24.72  $\pm$ 0.05 \\ \hline
GGAN \cite{haque2020guided} & $7.24 \pm 0.05$ & $25.75 \pm 0.10$ \\ \hline
GAAE & 7.28 $\pm$ 0.01 & 22.60 $\pm$ 0.07 \\ \hline
\end{tabular}
\end{table}

\begin{table}[t!]
\centering
\caption{{Comparison between the sample generation quality of the GAAE model and the other models for the Nsynth dataset. The generation quality is measured by IS and FID scores.}}

\label{tab:table2}
\begin{tabular}{|l|l|l|}\hline
\textbf{Model Name} & \textbf{IS Score} & \textbf{FID Score} \\ \hline
Real (Train Data) & 2.83 $\pm$ 0.02 &  -\\ \hline
Real (Test Data)  & 2.81 $\pm$ 0.12 & -\\ \hline
{\bf Supervised BigGAN} & {\bf 2.64 $\pm$ 0.08} & {\bf 148.30  $\pm$ 0.23}  \\ \hline
Unsupervised BigGAN & 2.21 $\pm$ 0.11 & 172.01  $\pm$ 0.15 \\ \hline
GGAN & 2.52 $\pm$ 0.06 & 149.23  $\pm$ 0.09 \\ \hline
GAAE & 2.58 $\pm$ 0.03 & 141.71 $\pm$ 0.32 \\ \hline
\end{tabular}
\end{table}

\begin{table*}[t!]
\centering
\caption{The relationship between the percentage of the data used as guidance during the training and the sample generation quality of the GAAE model, measured with the IS and the FID score. The scores are calculated for the S09 and the Nsynth dataset.}
\label{tab:table5}
\begin{tabular}{|l|l|l|l|l|}\hline
\textbf{Labelled Data} & \textbf{IS Score (S09)} & \textbf{FID Score (S09)}& \textbf{IS Score (Nsynth)} & \textbf{FID Score(Nsynth)} \\ \hline
1\%   & 6.94 $\pm$ 0.04 &  24.21 $\pm$ 0.16 & 2.48 $\pm$ 0.08 & 145.89 $\pm$ 1.32\\ \hline
2\%   & 7.06 $\pm$ 0.03 &  23.89 $\pm$ 0.11 & 2.53 $\pm$ 0.07 & 144.21 $\pm$ 0.65\\ \hline
3\%   & 7.12 $\pm$ 0.04 &  23.15 $\pm$ 0.10 & 2.56 $\pm$ 0.05 & 143.01 $\pm$ 0.43\\ \hline
4\%  & 7.19$\pm$ 0.02 &  22.91 $\pm$ 0.08 & 2.57 $\pm$ 0.04 & 142.46 $\pm$ 0.38\\ \hline
5\%   & 7.28 $\pm$ 0.01 &  22.60 $\pm$ 0.07 & 2.58 $\pm$ 0.03 & 141.71 $\pm$ 0.32 \\ \hline
{\bf 100\%}   & {\bf 7.45 $\pm$ 0.03} & {\bf 19.31 $\pm$ 0.01} & {\bf 2.67 $\pm$ 0.02} & {\bf 137.65 $\pm$ 0.02}\\ \hline

\end{tabular}
\end{table*}

%%%%%%%%%%%%%%%%%%%%%%%%%%%%TABLE%%%%%%%%%%%%%%%%%%%%%%%%%%%%%%%%%%%

%%%%%%%%%%%%%%%%%%%%%%%%%%%%FIG%%%%%%%%%%%%%%%%%%%%%%%%%%%%%%%%%%%

\Figure[!t]()[width=0.95\textwidth]{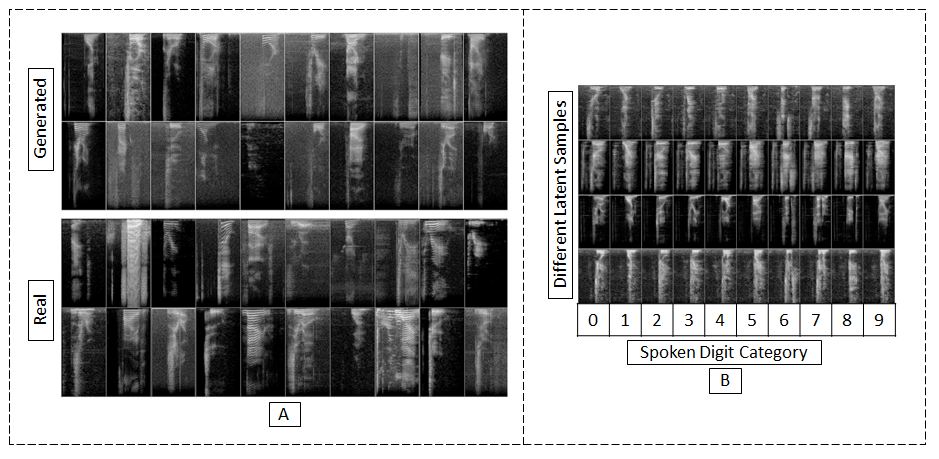}
   {A. Illustration of the difference between the generated spectrograms and the real spectrograms of the data for the S09 dataset. The top two rows show the randomly generated samples from the GAAE model, and the bottom two rows are the real samples from the training data. Notice the visual similarity between the generated and the real samples. B. This figure shows the generated spectrograms of the S09 dataset from the GAAE model according to different digit categories. Each row represents the samples generated for a fixed latent variable where the digit condition is changed from 0 to 9. Furthermore, any column shows the generated spectrogram for a particular digit category.\label{fig:real_generated_samples}}

\Figure[!t]()[width=0.90\textwidth]{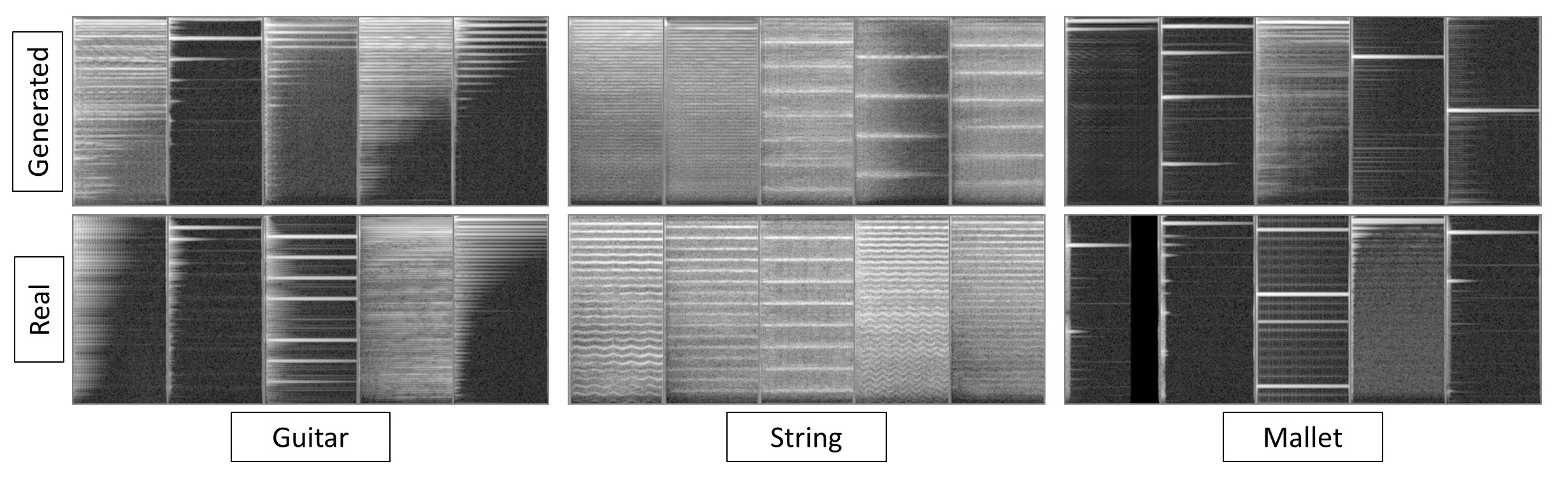}
   {Difference between the generated spectrograms of the GAAE model and the real spectrograms of the data for the Nsynth dataset. The top row shows the generated samples, and the bottom row shows the real samples. The first block shows the spectrogram of the guitar, and the other two illustrate the spectrograms for the strings and mallet.\label{fig:nsyth_all_samples}}

%%%%%%%%%%%%%%%%%%%%%%%%%%%%FIG%%%%%%%%%%%%%%%%%%%%%%%%%%%%%%%%%%%

\subsection{Evaluation of Conditional Sample Generation based on Guidance}
\subsubsection{Setup}

In this section, we evaluate the effectiveness of guidance for accurate conditional sample generation. It is cumbersome to check all the generation manually. Therefore, we manually check only a few audio samples. For large-scale validation, we use an approach similar to ~\cite{shmelkov}. We train a simple CNN classifier with the samples generated for different random conditions/categories and use the random categories associated with the generated samples as the true labels. Then, we evaluate the CNN classifier on the test dataset based on the classification accuracy. The rationale is that if the GAAE model does not learn to generate correct samples for any given category and the generated samples do not match the training data distribution; the CNN model will not be able to achieve good accuracy on the test dataset. We compare this CNN classifier with another CNN classifier which is trained using all the available training data. For further comparisons, we train two more CNN models with the generated samples from the supervised BigGAN and the GGAN model.

\subsubsection{Result and Discussion: Manual test}
The generated samples for the S09 and Nsynth dataset are shown in figure  \ref{fig:real_generated_samples} and figure  \ref{fig:nsyth_all_samples}, respectively. It is not visually evident that the model was able to generate correct samples according to the given conditions/categories. However, when we convert these spectrograms to audios, it is clear that the model is able to generate audios correctly according to the categories demonstrating the effectiveness of the guidance data to learn the specific categorical distribution of the training dataset (cf.\ under the above link).

\subsubsection{Results and Discussions: CNN based Classification accuracy}
For the S09 dataset, the test data classification accuracy for the CNN model trained with all the available labelled data is 95.52\% $\pm$ 0.50. The accuracy is 91.14\% $\pm$ 0.17, when the CNN model is trained based on the generated samples from the GAAE model (trained with 5\% labelled data). The table \ref{tab:table3} shows the comparison with other models. With the generated samples from the GAAE model, the CNN model achieves greater classification accuracy than the supervised BigGAN (86.58\% $\pm$ 0.56) and the GGAN model (86.72\% $\pm$ 0.47). 

When we trained the CNN model mixing the train data, and the generated samples from the GAAE model, the accuracy of the CNN model increased from  95.52\% $\pm$ 0.50 to 97.33\% $\pm$ 0.19. Along with the accuracy, the stability of the CNN model is also improved significantly.  This can be observed through the standard deviation in the results. We conducted the same evaluation on the Nsynth dataset and received similar results which we present in table \ref{tab:table4}. 

These results demonstrate the superior performance of our GAAE model for generating samples for different categories.
% With only 5\% labelled data as the guidance during training, the GAAE model learns superior conditional distribution of the training data and performs better than other models in terms of the sample diversity.
It can potentially be used as a data augmentation model where the generated samples from the model can be used to augment any related dataset or same dataset.

%%%%%%%%%%%%%%%%%%%%%%%%%%%%TABLE%%%%%%%%%%%%%%%%%%%%%%%%%%%%%%%%%%%

\begin{table}[t!]
\centering
\caption{The comparison between different CNN classifiers based on the test data classification accuracy from the S09 dataset. The CNN models are trained with the generated samples from different models.}
\label{tab:table3}
\begin{tabular}{|l|l|}\hline
\textbf{Sample for Training} & \textbf{Test Accuracy} \\ \hline
Train Data & 95.52\% $\pm$ 0.50\\\hline
Supervised BigGAN & 86.58\% $\pm$ 0.56\\ \hline
GGAN & 86.72\% $\pm$ 0.47\\ \hline
GAAE & 91.14\% $\pm$ 0.17 \\ \hline
{\bf GAAE + Train Data} & {\bf 97.33\% $\pm$ 0.19 } \\ \hline
\end{tabular}
\end{table}

\begin{table}[t!]
\centering
\caption{The comparison between different CNN classifiers based on the test data classification accuracy from the Nsynth dataset. The CNN models are trained with the generated samples from different models.}
\label{tab:table4}
\begin{tabular}{|l|l|}\hline
\textbf{Sample for Training} & \textbf{Test Accuracy} \\ \hline
Train Data & 92.01\% $\pm$ 0.94\\\hline
Supervised BigGAN & 83.50\% $\pm$ 0.62\\ \hline
GGAN & 81.40\% $\pm$ 0.48\\ \hline
GAAE & 86.80\% $\pm$ 0.23 \\ \hline
{\bf GAAE + Train Data} & {\bf 94.56\% $\pm$ 0.09} \\ \hline
\end{tabular}
\end{table}

%%%%%%%%%%%%%%%%%%%%%%%%%%%%TABLE%%%%%%%%%%%%%%%%%%%%%%%%%%%%%%%%%%%

\subsection{Conditional Sample Generation using guidance from a different dataset}

In the above two experiments, we used the guidance data from the same dataset. In this section, we explore the feasibility of guidance from a completely different dataset.

\subsubsection{Setup}
In the S09 dataset, there are both male and female speakers, but no label is available for the gender of the speakers.  We aim to verify if GAAE can generate samples from S09 dataset according to the condition on the gender category, where the guidance comes from a different dataset for gender category. To achieve this, we collect ten male and ten female speakers' audio data (randomly chosen with labels) from the Librispeech dataset to use as guidance during the training with the S09 dataset. During the training of the GAAE model, the guidance data from Librispeech dataset is also merged with S09 dataset as unlabelled data. So, GAAE learns to generate both samples from Librispeech dataset as well as from S09 dataset. 

The network we used before to calculate the IS and FID score, is trained on the digit classification tasks for S09 dataset, not for the gender classification task thus will no longer offer a meaningful evaluation. To eradicate this problem, we train another simple CNN model for the gender classification to calculate the IS and the FID score. For this purpose, we randomly select 15 male and 15 female speakers from Librispeech dataset. We use data from ten male and ten female speakers for training and data from others for testing. We achieve an accuracy of 98.3 $\pm$ 0.50. We use this model to calculate the IS and FID Score for the generated samples from different models. Now, the calculated scores will reflect the quality of the generated samples according to gender distribution. 

We define two GAAE models: one is trained with gender guidance, and another is trained with digit guidance. We compare the IS and FID score of these models. Note that gender information is being collected from a different dataset: Librispeech. If the gender guided model achieves better score, then we can establish the feasibility of guidance using an external dataset. To further validate this, we add results from other models (Unsupervised BigGAN, Supervised BigGAN and GGAN) trained based on digit guidance.

We choose a continuous normal distribution of size 128 for latent $z \sim \mathcal{N}(\mu = 0,\,\sigma^{2} = 1) $ and two gender categories for the conditions $y_{r} \sim Cat(y_{r}, K = 2,p = 0.5)$.

\subsubsection{Results and Discussions}
The calculated scores are presented in table \ref{tab:table6}. Gender guided GAAE produces the best FID and IS scores, which establish that it is feasible to get guidance from a different dataset in the GAAE model. 
%%%%%%%%%%%%%%%%%%%%%%%%%%%%TABLE%%%%%%%%%%%%%%%%%%%%%%%%%%%%%%%%%%%

\begin{table}[t!]
\centering
\caption{Comparison between the performance of the GGAN model trained with gender guidance and the other models on the S09 dataset, in terms of the quality of the generated samples based on the gender attributes of the speaker, measured with the IS and the FID score.}
\label{tab:table6}
\begin{tabular}{|l|l|l|}\hline
\textbf{Model Name} & \textbf{IS Score} & \textbf{FID Score} \\ \hline
Train Data  & 1.92 $\pm$ 0.04 & -\\\hline
Test Data  & 1.91 $\pm$ 0.05 & -\\ \hline
Unsupervised BigGAN & 1.13 $\pm$ 0.89 & 56.01 $\pm$ 0.85\\ \hline
Supervised BigGAN  & 1.48 $\pm$ 0.56 & 35.22 $\pm$ 0.50\\ \hline
GGAN (Digit Guided) & 1.58 $\pm$ 0.05 & 37.75 $\pm$ 0.10 \\\hline
GAAE (Digit Guided) & 1.61 $\pm$ 0.17 & 29.84 $\pm$ 0.43 \\\hline
{\bf GAAE (Gender Guided)} & {\bf 1.78 $\pm$ 0.03} & {\bf 20.21 $\pm$ 0.01} \\\hline

\end{tabular}
\end{table}

\subsection{Guided representation Learning}
The GAAE model learns two types of representations/latent spaces: (1) it uses $z_{x_{u}}$ $\sim$ $u_z$ to learn guidance specific characteristics of the data (Guided representation) and uses (2) $z^{'}_{x_{u}}$ $\sim$ $q_z$ to learn general characteristics of the data (General representation/Style representation).

\subsubsection{Setup}
In the GAAE model, the Classifier $C$ is built on top of the latent $z_{x_{u}}$ $\sim$ $u_z$ (see Fig.~\ref{fig:model_architecture}). The encoder network $E$, therefore, learns this latent variable to disentangle the class categories according to the guided data. For the S09 dataset, we use digit classes as guidance, so, in this latent space (representation space), the digit category should be disentangled. To observe this disentanglement, we visualise the higher dimensional (128) latent space generated for the S09 test data in the 2D plane with the t-SNE (t-distributed stochastic neighbour embedding) \cite{maaten2008visualizing} visualisation method. We use the same visualisation for the Nsynth dataset. 

\subsubsection{Results and Discussions}
Figure \ref{fig:S09_embedding} shows the representation space for S09 test dataset and figure \ref{fig:nsyth_embedding} shows the visualisation for the Nsynth dataset. From both figures, it is noticeable that the guided categories are well separated in the representation space, and data points of the similar categories are clustered together. So, the encoder $E$ learns to map the data sample to the representation space $u_z$ ensuring data categories used as guidance are well separable in the representation space.

%%%%%%%%%%%%%%%%%%%%%%%%%%%%FIG%%%%%%%%%%%%%%%%%%%%%%%%%%%%%%%%%%%

\Figure[!t]()[width=0.45\textwidth]{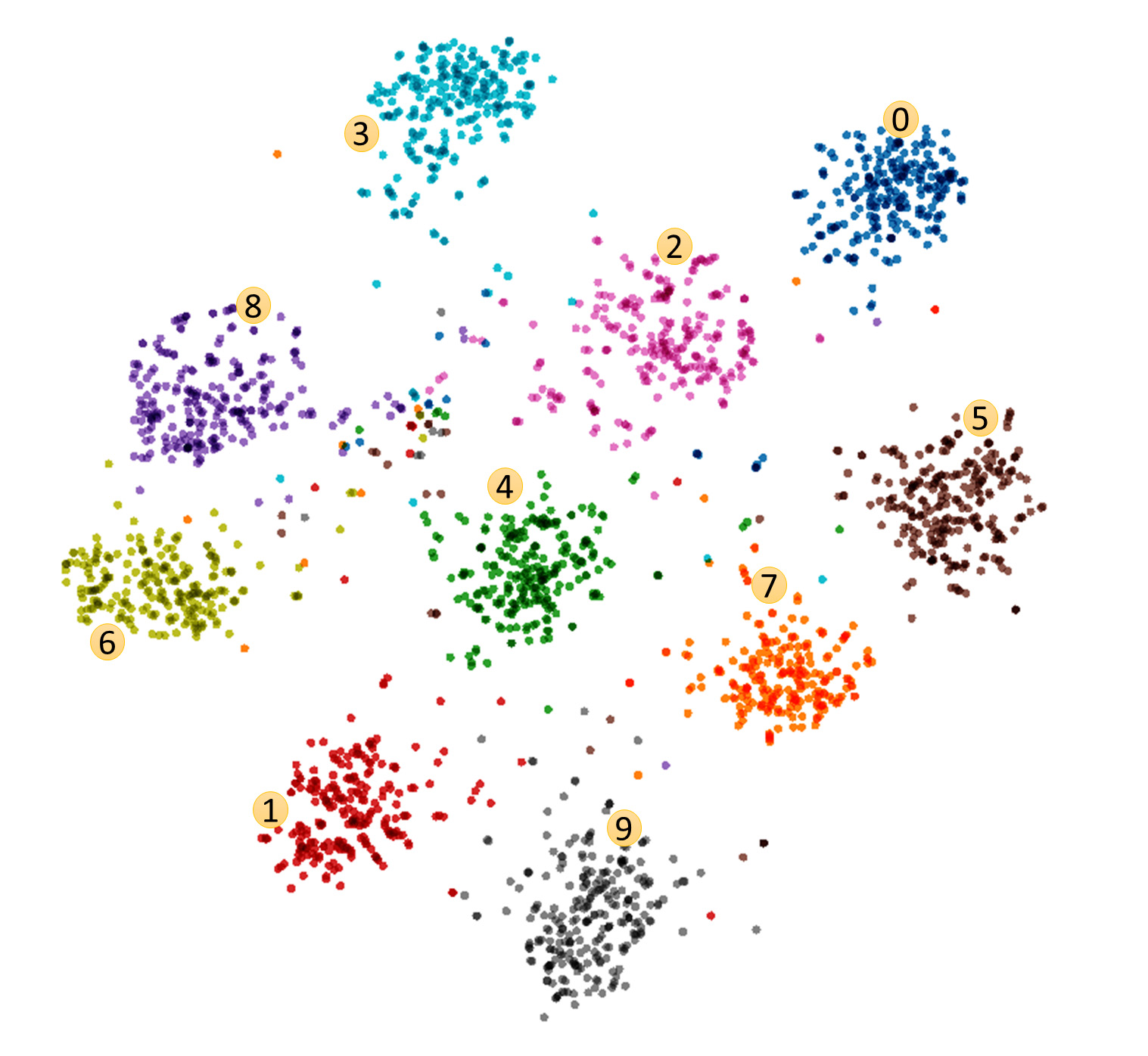}
   {t-SNE visualisation of the learnt representation of the test data of the S09 dataset. Here, different colours of points represent different digit categories. In the representation space, the different digit categories are clustered together and easily separable.\label{fig:S09_embedding}}
   
\Figure[!t]()[width=0.48\textwidth]{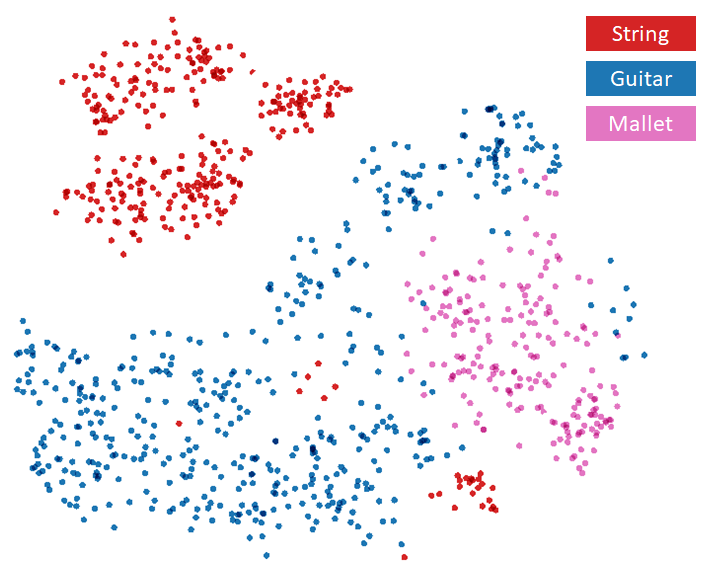}
   {t-SNE visualisation of the learnt representation of the test data of the Nsynth dataset. Here, different colours of points represent different instrument categories. In the representation space, the different instrument categories are clustered together and easily separable.\label{fig:nsyth_embedding}}
   
%%%%%%%%%%%%%%%%%%%%%%%%%%%%FIG%%%%%%%%%%%%%%%%%%%%%%%%%%%%%%%%%%%

\subsection{General Representation/Style Representation Learning}

\subsubsection{Setup}

The encoder network $E$ of the GAAE model is trained to match the $q_z$ distribution with the known $p_z$ distribution. This allows sampling $z^{'}_{x_{u}}$ from the $q_z$ distribution.

Now, it is expected that when Decoder $D$ learns to generate samples from the latent space $q_z$, it disentangles the general characteristics/attributes (independent of the guided attributes) of the data in the $q_z$ latent space. To evaluate this disentanglement in the representation space $z^{'}_{x_{u}}$ $\sim$ $q_z$ for both S09 and Nsynth dataset, we generate audio samples for different categories/conditions keeping the $z^{'}_{x_{u}}$ the same. 
 
In our model, Decoder can achieve disentanglement implies that the pretrained $E$ extracts general attributes in latent $z^{'}_{x_{u}}$ from any related dataset, which was not used during the training. To validate, we pass the test data from S09 and Nsynth dataset through $E$ to get the general representation $z^{'}_{x_{u}}$. Then for a fixed $z^{'}_{x_{u}}$ and different conditions (digit categories), we generate samples from the pretrained $D$ network.

As the GAAE model learns general/style attributes in the $z^{'}_{x_{u}}$ latent space, it should disentangle the gender of the speaker in the latent space for S09 dataset. To evaluate this, we use the trained $E$ network from the GAAE model to extract latent representation $z^{'}_{x_{u}}$ for an entirely different Librispeech dataset where gender labels are available. For 5000 randomly sampled data from the Librispeech dataset, we extract the feature/latent $z^{'}_{x_{u}}$ from $E$ and visualise the result in 2D plain using t-SNE visualisation for exploration.

\subsubsection{Results and Discussions}

After investigating the generated audios of the S09 dataset, the digit categories are changed according to the given condition $y_{r}$ and the general characteristics (such as the voice of the speaker, audio pitch, background noise etc.) of the audio is changed with the change of $z^{'}_{x_{u}}$. So, the $D$ network learns to capture general attributes of the data in the  latent space $z^{'}_{x_{u}}$. For the Nsynth dataset, we notice a similar behaviour.

We investigate the audio samples generated based on the extracted feature $z^{'}_{x_{u}}$ of the input data sample. Exploration of the audios shows that they preserve some characteristics (like speaker gender, voice, pitch, tone, background noise etc. for S09 test data) from the input data sample. We also notice similar scenarios for the Nsynth dataset. The audios can be found at: \href{https://bit.ly/36Oz9z9}{{https://bit.ly/36Oz9z9}}. Note that the initial one second is the input audio data and rest are the generated audios.

Figure \ref{fig:libri_gender} shows the visualisation of the extracted representation for the Librispeech dataset. We observe that the latent representation for the same gender of the speakers are clustered together and are easily separable from the latent space. This exploration exhibits that the GAAE model is able to learn the gender attributes of the speaker from the S09 dataset successfully even though gender information of the speaker was never used during the training.

%%%%%%%%%%%%%%%%%%%%%%%%%%%%FIG%%%%%%%%%%%%%%%%%%%%%%%%%%%%%%%%%%%

\Figure[!t]()[width=0.48\textwidth]{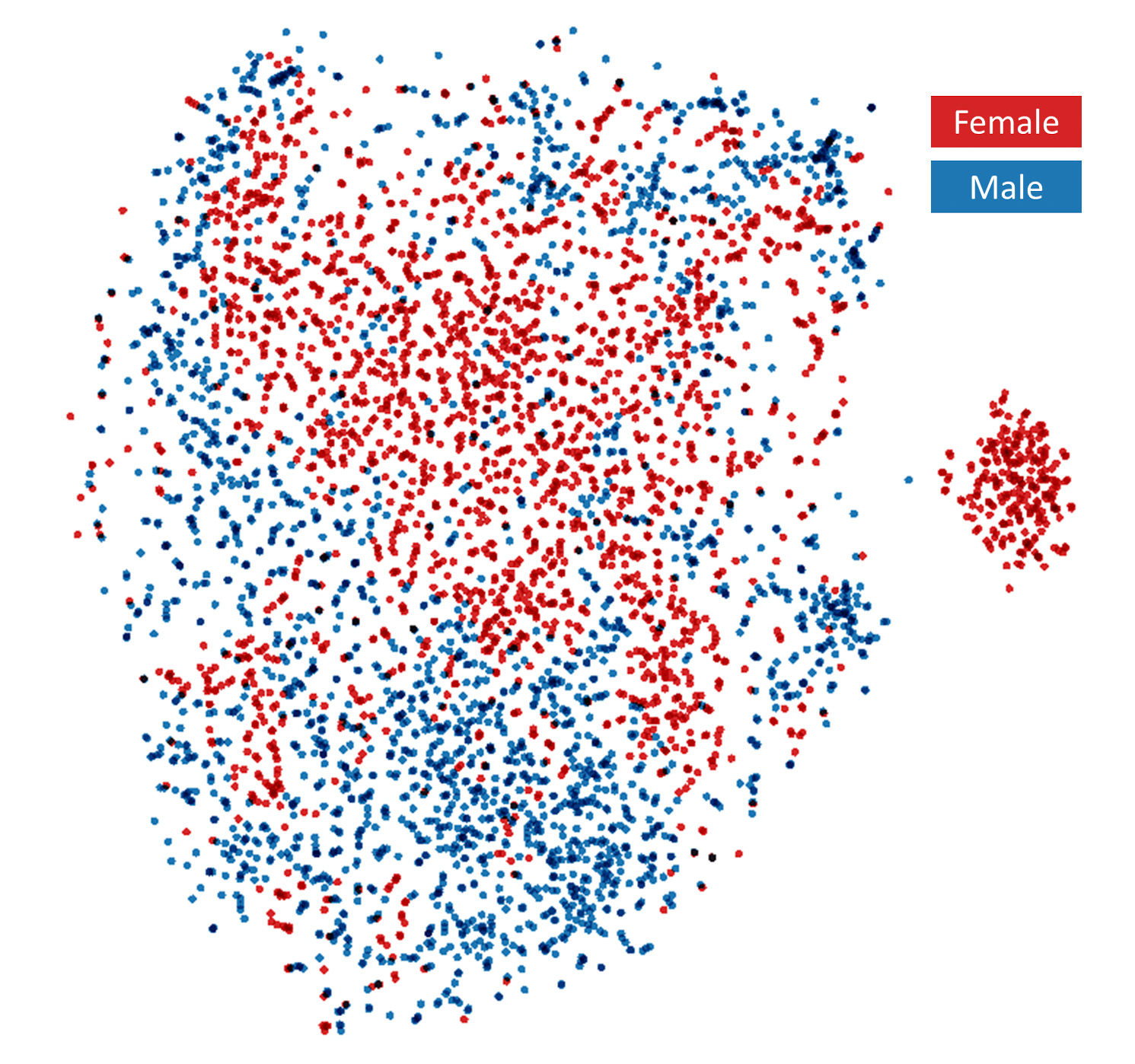}
   {t-SNE visualisation of the learnt representation of the Libri speech dataset. Here, different colours of points represent the gender of the speakers. The representations of the different gender categories are clustered together.\label{fig:libri_gender}}

%%%%%%%%%%%%%%%%%%%%%%%%%%%%FIG%%%%%%%%%%%%%%%%%%%%%%%%%%%%%%%%%%%

\subsection{Coherence of the General Representation/Latent Space}

\subsubsection{Setup}
It is expected that the $D$ network can learn the latent space $q_z$ in a way so that it is coherent and if we move in any direction in the latent space the generated samples should be changed accordingly. To investigate this, we conduct linear interpolation between two latent points as described in the DCGAN paper \cite{radford2015}. A particular point $z_{i}$ within two latent points $z_{0}$ and $z_{1}$ is calculated with the equation  $z_{i} = z_{0} + \eta (z_{0} - z_{1})$, where $\eta$ is the step size from $z_{0}$ to $z_{1}$. With this equation, we get the latent points in between $z_{0}$ and $z_{1}$. Using this $D$ network, we obtain the generated samples for these latent points, where the random categorical condition $y_{r}$ is fixed.

\subsubsection{Results}
Figure \ref{fig:linear_interpolation} shows the generated samples for both the S09 and Nsynth datasets based on the interpolated points. We observe that the transition between two spectrograms generated based on two fixed latent samples $z_{0}$ and $z_{1}$ is very smooth. Moreover, when we convert the spectrograms to audio, we observe the same smooth transition, which indicates the disentanglement of the general attributes in the latent space $q_{z}$ The audios can be found at: \href{https://bit.ly/2yPcTIE}{{https://bit.ly/2yPcTIE}}.

% Therefore, it appears evident from these explorations that the GAAE model is able to learn a  guided Representation as well as the representation for the general attributes/characteristics of the dataset.
%%%%%%%%%%%%%%%%%%%%%%%%%%%%FIG%%%%%%%%%%%%%%%%%%%%%%%%%%%%%%%%%%%

\Figure[!t]()[width=0.48\textwidth]{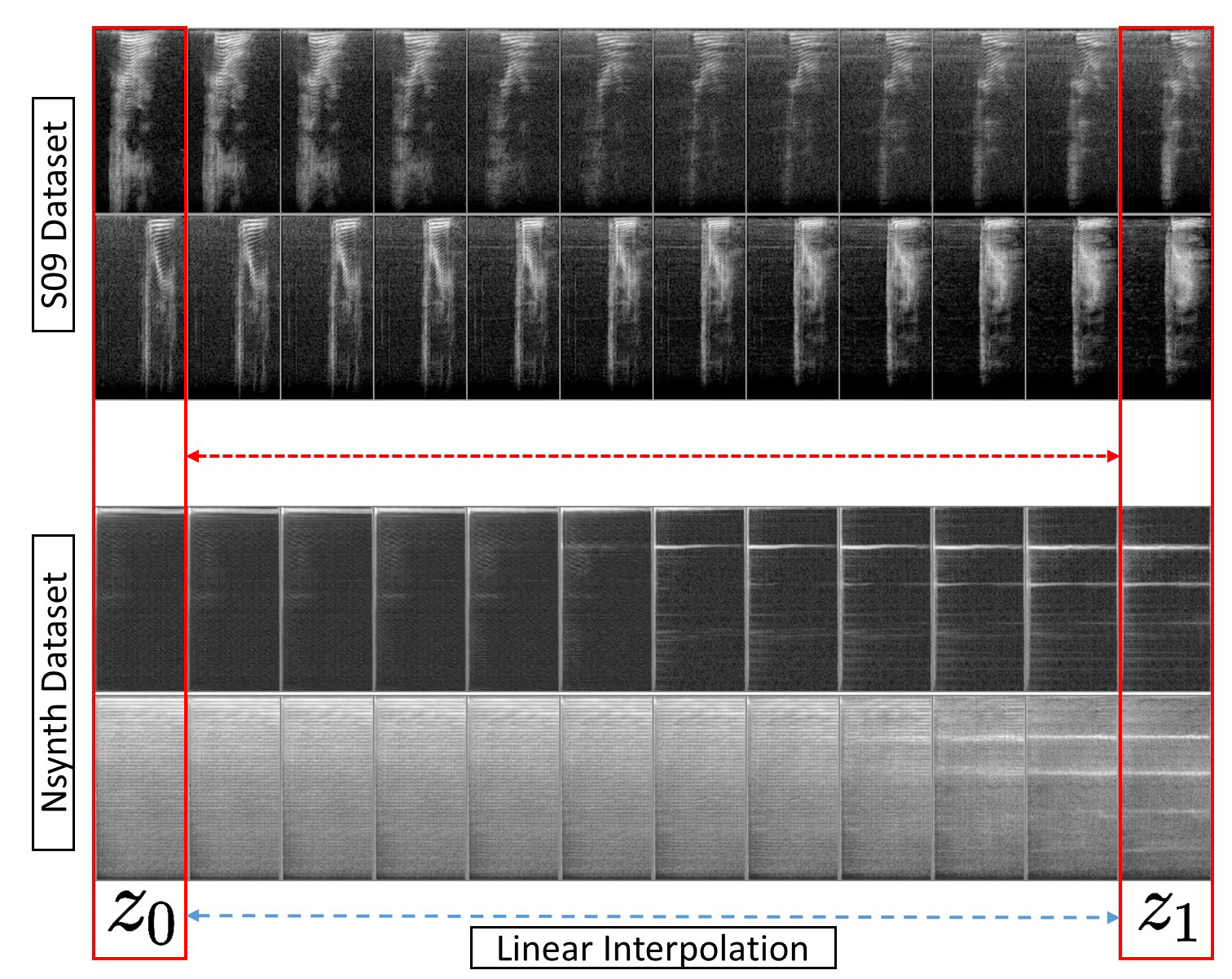}
   {Generated spectrograms based on the linear interpolation between two latent samples; $z_{0}$ and $z_{1}$. The first two rows show the generated spectrograms for the S09 dataset and the bottom two rows exhibit the spectrograms for the Nsynth dataset. For any particular row, the first and the last spectrograms are the generations based on the fixed two latent points and the in-between spectrograms are the generation based on the interpolation between these two fixed points.\label{fig:linear_interpolation}}

%%%%%%%%%%%%%%%%%%%%%%%%%%%%FIG%%%%%%%%%%%%%%%%%%%%%%%%%%%%%%%%%%%

\section{Hyperparameter Tuning}

We tune the hyperparameters based on the S09 dataset as tuning is resource and time-intensive. We then use the hyperparameters for other datasets. From equation \ref{eq:5}, $\omega_{1}$, and $\omega_{2}$ are two important hyperparameters for training the GAAE model, where $\omega_{2}$ = 1 - $\omega_{1}$. When we increase  $\omega_{1}$, the model focuses more on the generation loss $G_{loss}$ and less on the reconstruction loss $R_{loss}$. If we reduce $\omega_{1}$, the model increases the focus for reconstruction and reduces the focus for the generation. The impact of $\omega_{1}$ and $\omega_{2}$ on the IS scores, FID scores, and classification accuracy are presented in figure \ref{fig:alpha}. The best value for  $\omega_{1}$ is 0.6 and for $\omega_{2}$, it is 0.4. 

The $\alpha$ and $\beta$ from equation \ref{eq:5} are two other important hyperparameters. The value of the $\alpha$ parameter determines how much the model will focus on generation ($G_{loss}$) and reconstruction loss ($R_{loss}$), where the $\beta$ parameter determines the focus for the classification ($Cl_{loss}$, $Cg_{loss}$) and latent generation loss ($L_{loss}$). From figure \ref{fig:alpha}, we observe that 0.5 is the best value for both of the hyperparameters. 

There are three more hyperparameters:  $\omega_{3}$, $\omega_{4}$, and $\omega_{5}$ (See equation \ref{eq:5}). Here, $\omega_{3}$ and $\omega_{4}$ control the classification loss ($Cl_{loss}$, $Cg_{loss}$) for labelled data. And, $\omega_{5}$ controls the latent generation loss ($L_{loss}$). Here, we maintain equal balance between the classification and the latent generation loss. Likewise, we use 0.25 for $\omega_{3}$,$\omega_{4}$ and 0.50 for $\omega_{5}$.

%%%%%%%%%%%%%%%%%%%%%%%%%%%%FIG%%%%%%%%%%%%%%%%%%%%%%%%%%%%%%%%%%%
\Figure[!t]()[width=0.95\textwidth]{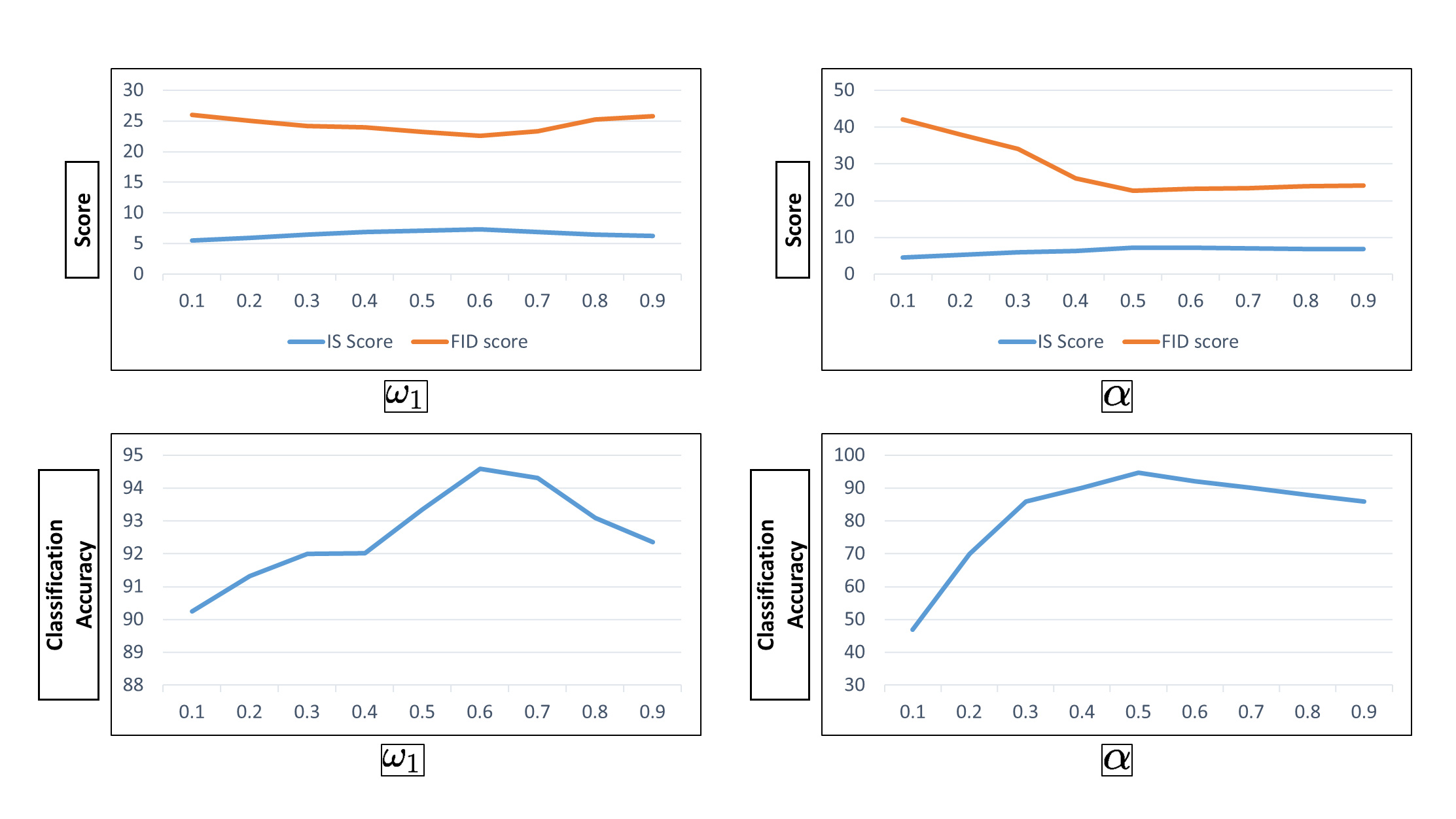}
   {Relationship between the hyperparameters and the measurement metrics of the GAAE model. The top left plot explains the relationship between  $\omega_{1}$ and IS and FID scores. Similarly, the top right explicates the relationship between $\alpha$ and IS and FID scores. Here, The bottom left box illustrates the relationship between $\omega_{1}$ and the classification accuracy. Furthermore, the bottom right plot demonstrates the impact of $\alpha$ on the classification accuracy.\label{fig:alpha}}
%%%%%%%%%%%%%%%%%%%%%%%%%%%%FIG%%%%%%%%%%%%%%%%%%%%%%%%%%%%%%%%%%%

% We tuned the hyperparameters based on the S09 dataset
% only because the tuning costs extensive amount of resource
% and time. Then, we used those hyperparameters for other
% datasets. From equation 6, ω1 and ω2 are two important
% hyperparameters for training the GAAE model, where ω2 = 1
% - ω1. When we increase the ω1, the model focuses more on the
% generation and less on the reconstruction. Now if we reduce
% the ω1, the model increases the focus for reconstruction and
% reduces the focus for the generation. The relationship between
% ω1 and the IS score, FID score, Classification accuracy can
% be found in figure 9. The optimal value for the ω1 is 0.6
% and for the ω2 it is 0.4. The hyperparameters α and β from
% the equation 6 are two main hyperparameters. The value
% of α parameter determines how much the model will focus
% on generation and reconstruction loss where β parameter
% determines the focus for the classification and latent loss.
% From figure 9, we can observe that 0.5 is the optimal value
% for both of the hyperparameters. In the equation 6 there are
% three more hyperparameters; ω1,ω2 and ω3. Hence, ω1 and ω2
% determines the focus for classification loss for labelled data,
% generated data respectively, where the ω3 determines the focus
% for the latent loss. Here, equal balance is optimal between the
% classification and latent loss. So we have used 0.25 for ω1,ω2
% and 0.50 for the ω3.

%%%%%%%%%%%%%%%%%%%%%%%%%%%%Table%%%%%%%%%%%%%%%%%%%%%%%%%%%%%%%%%%%

\begin{table*}[t!]
\centering
\caption{Relationship between the percentage of the data used as the guidance during the training and the S09 test dataset classification accuracy of the GAAE model.}
\label{tab:table7}
\begin{tabular}{|l|l|l|l|l|}\hline
\textbf{\begin{tabular}[c]{@{}l@{}}Training\\ Data Size\end{tabular}} &
\textbf{\begin{tabular}[c]{@{}l@{}}CNN\\ Network\end{tabular}} & \textbf{\begin{tabular}[c]{@{}l@{}}BiGAN\end{tabular}} & \textbf{GGAN}  & \textbf{GAAE} \\\hline
1\%& 82.21 $\pm$  1.2 & 73.01 $\pm$  1.02 &  84.21 $\pm$  2.24&  90.21 $\pm$  0.16\\ \hline
2\% & 83.04  $\pm$ 0.34 & 75.56 $\pm$  0.41 &  85.39 $\pm$  1.24&  91.45 $\pm$  0.12\\ \hline
3\% & 83.78  $\pm$ 0.23 & 78.33 $\pm$  0.07 & 88.25 $\pm$ 0.10 &  92.67 $\pm$  0.06\\ \hline
4\% & 84.11  $\pm$ 0.34 & 80.03 $\pm$  0.01 &  91.02 $\pm$  0.50&  93.70 $\pm$  0.05 \\ \hline
5\% & 84.50 $\pm$  1.02 & 80.84 $\pm$  1.72 &  92.00  $\pm$ 0.87&  94.59 $\pm$  0.03\\ \hline
{\bf 100\%} & {\bf 95.52 $\pm$ 0.50} & {\bf 86.77 $\pm$  2.61} &  {\bf 96.51$\pm$ 0.07}  &  {\bf 97.68 $\pm$ 0.01} \\ \hline
\end{tabular}
\end{table*}

\begin{table*}[t!]
\centering
\caption{Relationship between the percentage of the data used as the guidance during the training and the Nsynth test dataset classification accuracy of the GAAE model.}
\label{tab:table8}
\begin{tabular}{|l|l|l|l|l|}\hline
\textbf{\begin{tabular}[c]{@{}l@{}}Training\\ Data Size\end{tabular}} &
\textbf{\begin{tabular}[c]{@{}l@{}}CNN\\ Network\end{tabular}} & \textbf{\begin{tabular}[c]{@{}l@{}}BiGAN\end{tabular}} & \textbf{GGAN}  & \textbf{GAAE} \\\hline
1\%& 85.76 $\pm$  1.10 & 82.21 $\pm$ 0.84 & 88.52 $\pm$ 0.32 & 90.26 $\pm$ 0.09\\ \hline
2\% & 89.79  $\pm$ 0.51 & 86.65 $\pm$ 0.57 & 91.69 $\pm$ 0.24 & 92.96 $\pm$ 0.07\\ \hline
3\% & 89.83  $\pm$ 0.49 & 87.21 $\pm$ 0.46 & 91.95 $\pm$ 0.20 & 93.12 $\pm$ 0.05\\ \hline
4\% & 90.52  $\pm$ 0.25 & 87.59 $\pm$ 0.41 & 92.16 $\pm$ 0.19 & 93.73 $\pm$ 0.02\\ \hline
5\% & 91.07 $\pm$  0.31 & 87.95 $\pm$ 0.39 & 92.45 $\pm$ 0.14 & 94.23 $\pm$ 0.02\\ \hline
{\bf 100\%} & {\bf 92.01 $\pm$  0.94} & {\bf 88.09 $\pm$ 0.24} & {\bf 93.56 $\pm$ 0.09 } & {\bf 94.89 $\pm$ 0.01} \\ \hline
\end{tabular}
\end{table*}

%%%%%%%%%%%%%%%%%%%%%%%%%%%%Table%%%%%%%%%%%%%%%%%%%%%%%%%%%%%%%%%%%

\section{Classifier of the GAAE model}

The success of the GAAE model is mostly dependent on its internal Classifier $C$. In this section, we evaluate the performance of $C$. We benchmark its performance using a supervised Classifier, the Classifier from GGAN and the Classifier from BiGAN \cite{DonahueKD16}. For the supervised Classifier, we train a simple CNN classifier using 1\,\% - 5\,\%, 100\,\%, of training data, where the data is heavily augmented using techniques like adding random noise, rotation of the spectrogram, multiplication with random zero patches, etc. (\cite{park2019specaugment}). We train a BiGAN model on top of the unsupervised BigGAN and extract BiGANs' feature network after the training. We then train another feed-forward classifier network on BiGANs' feature network using similar percentages of labelled data. We keep the weights for the feature network fixed during the training. We evaluate all these Classifiers using the test dataset. As the Classifier $C$ of the GAAE model is trained with fewer labelled data along with the generated samples from the decoder $D$, it will only perform better if generation is accurate according to the different categories and the quality of the generated samples is close to the real samples.

The relationship between the percentage of the data used as guidance and the test data classification accuracy is shown in table \ref{tab:table7}, \ref{tab:table8} for S09 and Nsyth dataset, respectively.  
Results from both tables demonstrate that the GAEE model outperforms other models in terms of classification accuracy leveraging the minimal amount of labelled data.

\section{Conclusion and Lesson Learnt}

In this paper, we propose the Guided Adversarial Autoencoder (GAAE), which is capable of generating high-quality audio samples using very few labelled data as guidance. After evaluating the GAAE model using two audio datasets: S09 and Nsynth, we show that the GAAE model can outperform the existing models with respect to sample generation quality and mode diversity. Harnessing the power of high-fidelity audio generation, the GAAE model can disentangle the specific attributes of the data in the learnt latent/representation space according to the guidance. This learnt representation can be beneficial to any related downstream task at hand. We also show that besides the guided representation learning, the GAAE model learns to disentangle other attributes of the data independent of the given guidance. Hence, the GAAE model learns a representation for the specific downstream task at hand and a generalised representation for future unknown related tasks.  

We evaluate the GAAE model based on the audio of size one second; thus, it remains a challenge to make this model work for longer audio sample generation. In representation learning, the GAAE model can be used efficiently for any long audio sample by dividing it into one-second chunks. GAAE model successfully learns generation and representation using a minimum of 1\,\% labelled data. We believe this will encourage other researchers to explore the GAAE model further for few-shot learning. 

Furthermore, we built the GAAE model based on BigGAN architecture. This leaves an excellent opportunity for studying other high performing GAN architectures such as progressive GAN\cite{karras:2017} or the Style GAN\cite{karras2019style}.

\newpage %BS finished editing

\medskip
\newpage
\bibliographystyle{ieeetr}
\bibliography{references}

\appendices

\section{ARCHITECTURAL DETAILS}
This section presents the details of the neural networks used in this paper. We follow the abbreviations and description style from the original work of Mario et al.\ \cite{lucic2019highfidelity}.

\begin{table}[h!]
\centering
\caption{Abbreviations for defining the architectures.}
\begin{tabular}{|l|l|}
\hline
\textbf{\begin{tabular}[c]{@{}l@{}} Full Name\end{tabular}} & \textbf{\begin{tabular}[c]{@{}l@{}}Abbreviation\\ \end{tabular}} \\ \hline
Resample & RS \\ \hline
Batch normalisation & BN \\ \hline
Conditional batch normalisation & cBN \\ \hline
Downscale & D \\ \hline
Upscale & U \\ \hline
%None & - \\ \hline
Spectral normalisation & SN \\ \hline
Input height & h \\ \hline
Input width & w \\ \hline
True label & y \\ \hline
Input channels & ci \\ \hline
Output channels & co \\ \hline
Number of channels  & ch \\ \hline
\end{tabular}
\label{tab:tab0}
\end{table}

\subsection{Supervised BigGAN}

We use the exact implementation of the Supervised BigGAN from our former GGAN paper \cite{haque2020guided}. Therefore, for the implementation of both the Generator and the  Discriminator, we apply a Resnet architecture from the BigGAN work\cite{Andrew_biggan}. The layers are shown tables \ref{tab:tab1} and \ref{tab:tab2}. The Generator and Discriminator architectures are shown in Tables \ref{tab:tab3} and \ref{tab:tab4}, respectively. 
We use a learning rate of $0.00005$ and $0.0002$ for the Generator and the Discriminator, respectively. We set the number of channels (ch) to 16 to minimise the computational expenses, as the higher number of channels such as 64 and 32 only offer negligible improvements.

\begin{table}[h!]
\centering
\caption{Architecture of the ResBlock generator with upsampling for the supervised BigGAN.}
\begin{tabular}{|l|l|l|l|}\hline
\textbf{\begin{tabular}[c]{@{}l@{}}Layer\\ Name\end{tabular}} & \textbf{\begin{tabular}[c]{@{}l@{}}Kernal\\ Size\end{tabular}} & \textbf{RS} & \textbf{\begin{tabular}[c]{@{}l@{}}Output\\ Size\end{tabular}} \\ \hline
Shortcut & {[}1,1,1{]} & U & 2h $\times$ 2w $\times$ c\_\{o\} \\ \hline
cBN, ReLU & - & - & h $\times$ w $\times$ c\_\{i\} \\ \hline
Convolution & {[}3,3,1{]} & U & 2h $\times$ 2w $\times$ c\_\{o\} \\ \hline
cBN, ReLU & - & - & 2h $\times$ 2w $\times$ c\_\{o\} \\ \hline
Convolution & {[}3,3,1{]} & U & 2h $\times$ 2w $\times$ c\_\{o\} \\ \hline
Addition & - & - & 2h $\times$ 2w $\times$ c\_\{o\} \\ \hline

\end{tabular}
\label{tab:tab1}
\end{table}

\begin{table}[h!]
\centering
\caption{Architecture of the ResBlock discriminator with downsampling for the supervised BigGAN.}
\begin{tabular}{|l|l|l|l|}
\hline
\textbf{\begin{tabular}[c]{@{}l@{}}Layer\\ Name\end{tabular}} & \textbf{\begin{tabular}[c]{@{}l@{}}Kernal\\ Size\end{tabular}} & \textbf{RS} & \textbf{\begin{tabular}[c]{@{}l@{}}Output\\ Size\end{tabular}} \\ \hline
Shortcut & {[}1,1,1{]} & D & h/2 $\times$ w/2 $\times$ c\_\{o\} \\ \hline
ReLU & - & - & h $\times$ w $\times$ c\_\{i\} \\ \hline
Convolution & {[}3,3,1{]} & - & h $\times$ w $\times$ c\_\{o\} \\ \hline
ReLU & - & - & h $\times$ w $\times$ c\_\{o\} \\ \hline
Convolution & {[}3,3,1{]} & D & h/2 $\times$ w/2 $\times$ c\_\{o\} \\ \hline
Addition & - & - & h/2 $\times$ w/2 $\times$ c\_\{o\} \\ \hline

\end{tabular}
\label{tab:tab2}
\end{table}

% "ch" refers to channel wide multiplier.we have used 16 for "ch".

\begin{table}[h!]
\centering
\caption{Architecture of the generator for the supervised BigGAN.}
\begin{tabular}{|l|l|l|l|}
\hline
\textbf{\begin{tabular}[c]{@{}l@{}}Layer\\ Name\end{tabular}} & \textbf{RS} & \textbf{SN} & \textbf{\begin{tabular}[c]{@{}l@{}}Output\\ Size\end{tabular}} \\ \hline
Input z & - & - & 128 \\ \hline
Dense & - & - & 4 $\times$ 2 $\times$ 16. ch \\ \hline
ResBlock & U & SN & 8 $\times$ 4 $\times$ 16. ch \\ \hline
ResBlock & U & SN & 16 $\times$ 8 $\times$ 16. ch \\ \hline
ResBlock & U & SN & 32 $\times$ 16 $\times$ 16. ch \\ \hline
ResBlock & U & SN & 64 $\times$ 32 $\times$ 16. ch \\ \hline
ResBlock & U & SN & 128 $\times$ 64 $\times$ 16. ch \\ \hline
Non-local block & - & - & 128 $\times$ 64 $\times$ 16. ch \\ \hline
ResBlock & U & SN & 256 $\times$ 128 $\times$ 1. ch \\ \hline
BN, ReLU & - & - & 256 $\times$ 128 $\times$ 1 \\ \hline
Conv {[}3, 3, 1{]} & - & - & 256 $\times$ 128 $\times$ 1 \\ \hline
Tanh & - & - & 256 $\times$ 128 $\times$ 1 \\ \hline

\end{tabular}
\label{tab:tab3}
\end{table}

\begin{table}[h!]
\centering
\caption{Architecture of the discriminator for the supervised BigGAN.}
\begin{tabular}{|l|l|l|}
\hline
\textbf{\begin{tabular}[c]{@{}l@{}}Layer\\ Name\end{tabular}} & \textbf{RS} & \textbf{\begin{tabular}[c]{@{}l@{}}Output\\ Size\end{tabular}} \\ \hline
\begin{tabular}[c]{@{}l@{}}Input \\ Spectrogram\end{tabular} & - & 256 $\times$ 128 $\times$ 1 \\ \hline
ResBlock & D & 128 $\times$ 64 $\times$ 1. ch \\ \hline
Non-local block & - & 128 $\times$ 64 $\times$ 1. ch \\ \hline
ResBlock & - & 64 $\times$ 32 $\times$ 1. ch \\ \hline
ResBlock & D & 32 $\times$ 16 $\times$ 2. ch \\ \hline
ResBlock & D & 16 $\times$ 8 $\times$ 4. ch \\ \hline
ResBlock & D & 8 $\times$ 4 $\times$ 8. ch \\ \hline
ResBlock & D & 4 $\times$ 2 $\times$ 16. ch \\ \hline
\begin{tabular}[c]{@{}l@{}}ResBlock \\ (No Shortcut)\end{tabular} & - & 4 $\times$ 2 $\times$ 16. ch \\ \hline
ReLU & - & 4 $\times$ 2 $\times$ 16. ch \\ \hline
Global sum pooling & - & 1 $\times$ 1 $\times$ 16. ch \\ \hline
Sum(embed(y)·h)+(dense $\rightarrow$ 1) & - & 1 \\ \hline

\end{tabular}
\label{tab:tab4}
\end{table}

\subsection{Unsupervised BigGAN}
Similarly, for the unsupervised BigGAN, follow the same implementation from the original GGAN work \cite{haque2020guided}. Tables \ref{tab:tab5} and \ref{tab:tab6} show the  upsampling and downsampling layers, respectively.  The architectures of the Generator and Discriminator are shown in the tables \ref{tab:tab7} and \ref{tab:tab8}, respectively. The learning rate and channels are the same as for the supervised BigGAN. 

\begin{table}[h!]
\centering
\caption{Architecture of the ResBlock generator with upsampling for the unsupervised BigGAN.}

\begin{tabular}{|l|l|l|l|}
\hline
\textbf{\begin{tabular}[c]{@{}l@{}}Layer\\ Name\end{tabular}} & \textbf{\begin{tabular}[c]{@{}l@{}}Kernal\\ Size\end{tabular}} & \textbf{RS} & \textbf{\begin{tabular}[c]{@{}l@{}}Output\\ Size\end{tabular}} \\ \hline
Shortcut & {[}1,1,1{]} & U & 2h $\times$ 2w $\times$ c\_\{o\} \\ \hline
BN, ReLU & - & - & h $\times$ w $\times$ c\_\{i\} \\ \hline
Convolution & {[}3,3,1{]} & U & 2h $\times$ 2w $\times$ c\_\{o\} \\ \hline
BN, ReLU & - & - & 2h $\times$ 2w $\times$ c\_\{o\} \\ \hline
Convolution & {[}3,3,1{]} & U & 2h $\times$ 2w $\times$ c\_\{o\} \\ \hline
Addition & - & - & 2h $\times$ 2w $\times$ c\_\{o\} \\ \hline

\end{tabular}
\label{tab:tab5}
\end{table}

\begin{table}[h!]
\centering
\caption{Architecture of the ResBlock discriminator with downsampling for the unsupervised BigGAN.}
\begin{tabular}{|l|l|l|l|}
\hline
\textbf{\begin{tabular}[c]{@{}l@{}}Layer\\ Name\end{tabular}} & \textbf{\begin{tabular}[c]{@{}l@{}}Kernal\\ Size\end{tabular}} & \textbf{RS} & \textbf{\begin{tabular}[c]{@{}l@{}}Output\\ Size\end{tabular}} \\ \hline
Shortcut & {[}1,1,1{]} & D & h/2 $\times$ w/2 $\times$ c\_\{o\} \\ \hline
ReLU & - & - & h $\times$ w $\times$ c\_\{i\} \\ \hline
Convolution & {[}3,3,1{]} & - & h $\times$ w $\times$ c\_\{o\} \\ \hline
ReLU & - & - & h $\times$ w $\times$ c\_\{o\} \\ \hline
Convolution & {[}3,3,1{]} & D & h/2 $\times$ w/2 $\times$ c\_\{o\} \\ \hline
Addition & - & - & h/2 $\times$ w/2 $\times$ c\_\{o\} \\ \hline

\end{tabular}
\label{tab:tab6}
\end{table}

% "ch" refers to channel wide multiplier.we have used 16 for "ch".

\begin{table}[h!]
\centering
\caption{Architecture of the generator for the unsupervised BigGAN.}
\begin{tabular}{|l|l|l|l|}
\hline
\textbf{\begin{tabular}[c]{@{}l@{}}Layer\\ Name\end{tabular}} & \textbf{RS} & \textbf{SN} & \textbf{\begin{tabular}[c]{@{}l@{}}Output\\ Size\end{tabular}} \\ \hline
Input z & - & - & 128 \\ \hline
Dense & - & - & 4 $\times$ 2 $\times$ 16. ch \\ \hline
ResBlock & U & SN & 8 $\times$ 4 $\times$ 16. ch \\ \hline
ResBlock & U & SN & 16 $\times$ 8 $\times$ 16. ch \\ \hline
ResBlock & U & SN & 32 $\times$ 16 $\times$ 16. ch \\ \hline
ResBlock & U & SN & 64 $\times$ 32 $\times$ 16. ch \\ \hline
ResBlock & U & SN & 128 $\times$ 64 $\times$ 16. ch \\ \hline
Non-local block & - & - & 128 $\times$ 64 $\times$ 16. ch \\ \hline
ResBlock & U & SN & 256 $\times$ 128 $\times$ 1. ch \\ \hline
BN, ReLU & - & - & 256 $\times$ 128 $\times$ 1 \\ \hline
Conv {[}3, 3, 1{]} & - & - & 256 $\times$ 128 $\times$ 1 \\ \hline
Tanh & - & - & 256 $\times$ 128 $\times$ 1 \\ \hline

\end{tabular}
\label{tab:tab7}
\end{table}

\begin{table}[h!]
\centering
\caption{Architecture of the discriminator for the unsupervised BigGAN.}
\begin{tabular}{|l|l|l|}
\hline
\textbf{\begin{tabular}[c]{@{}l@{}}Layer\\ Name\end{tabular}} & \textbf{RS} & \textbf{\begin{tabular}[c]{@{}l@{}}Output\\ Size\end{tabular}} \\ \hline
\begin{tabular}[c]{@{}l@{}}Input \\ Spectrogram\end{tabular} & - & 256 $\times$ 128 $\times$ 1 \\ \hline
ResBlock & D & 128 $\times$ 64 $\times$ 1. ch \\ \hline
Non-local block & - & 128 $\times$ 64 $\times$ 1. ch \\ \hline
ResBlock & - & 64 $\times$ 32 $\times$ 1. ch \\ \hline
ResBlock & D & 32 $\times$ 16 $\times$ 2. ch \\ \hline
ResBlock & D & 16 $\times$ 8 $\times$ 4. ch \\ \hline
ResBlock & D & 8 $\times$ 4 $\times$ 8. ch \\ \hline
ResBlock & D & 4 $\times$ 2 $\times$ 16. ch \\ \hline
\begin{tabular}[c]{@{}l@{}}ResBlock \\ (No Shortcut)\end{tabular} & - & 4 $\times$ 2 $\times$ 16. ch \\ \hline
ReLU & - & 4 $\times$ 2 $\times$ 16. ch \\ \hline
Global sum pooling & - & 1 $\times$ 1 $\times$ 16. ch \\ \hline
Dense & - & 1 \\ \hline

\end{tabular}
\label{tab:tab8}
\end{table}

\subsection{BiGAN}

For the BiGAN model, we train a Feature Extractor and Discriminator network on top of the unsupervised BigGAN. The Feature Extractor network creates the features for real samples, and the Discriminator tries to differentiate between the generated features and the random noise. The detail is exactly followed from the original BiGAN work \cite{DonahueKD16}. 
The downsampling layer is the same as the unsupervised BigGAN and can be found in table \ref{tab:tab6}. The architecture of the Feature Extractor network is shown in table \ref{tab:tab9}. Furthermore, the architecture of the Discriminator is given in table \ref{tab:tab10}.

\begin{table}[ht!]
\centering
\caption{Architecture of the Feature Extractor Network for the BiGAN.}
\begin{tabular}{|l|l|l|}
\hline
\textbf{\begin{tabular}[c]{@{}l@{}}Layer\\ Name\end{tabular}} & \textbf{RS} & \textbf{\begin{tabular}[c]{@{}l@{}}Output\\ Size\end{tabular}} \\ \hline
\begin{tabular}[c]{@{}l@{}}Input \\ Spectrogram\end{tabular} & - & 256 $\times$ 128 $\times$ 1 \\ \hline
ResBlock & D & 128 $\times$ 64 $\times$ 1. ch \\ \hline
Non-local block & - & 128 $\times$ 64 $\times$ 1. ch \\ \hline
ResBlock & - & 64 $\times$ 32 $\times$ 1. ch \\ \hline
ResBlock & D & 32 $\times$ 16 $\times$ 2. ch \\ \hline
ResBlock & D & 16 $\times$ 8 $\times$ 4. ch \\ \hline
ResBlock & D & 8 $\times$ 4 $\times$ 8. ch \\ \hline
ResBlock & D & 4 $\times$ 2 $\times$ 16. ch \\ \hline
\begin{tabular}[c]{@{}l@{}}ResBlock \\ (No Shortcut)\end{tabular} & - & 4 $\times$ 2 $\times$ 16. ch \\ \hline
ReLU & - & 4 $\times$ 2 $\times$ 16. ch \\ \hline
Global sum pooling & - & 1 $\times$ 1 $\times$ 16. ch \\ \hline
Dense & - & 128 \\ \hline

\end{tabular}
\label{tab:tab9}
\end{table}

\begin{table}[ht!]
\centering
\caption{Architecture of the Discriminator for the BiGAN.}
\begin{tabular}{|l|l|l|}
\hline
\textbf{\begin{tabular}[c]{@{}l@{}}Layer\\ Name\end{tabular}} & \textbf{RS} & \textbf{\begin{tabular}[c]{@{}l@{}}Output\\ Size\end{tabular}} \\ \hline
\begin{tabular}[c]{@{}l@{}}Input \\ Spectrogram\end{tabular} & - & 256 $\times$ 128 $\times$ 1 \\ \hline
ResBlock & D & 128 $\times$ 64 $\times$ 1. ch \\ \hline
Non-local block & - & 128 $\times$ 64 $\times$ 1. ch \\ \hline
ResBlock & - & 64 $\times$ 32 $\times$ 1. ch \\ \hline
ResBlock & D & 32 $\times$ 16 $\times$ 2. ch \\ \hline
ResBlock & D & 16 $\times$ 8 $\times$ 4. ch \\ \hline
ResBlock & D & 8 $\times$ 4 $\times$ 8. ch \\ \hline
ResBlock & D & 4 $\times$ 2 $\times$ 16. ch \\ \hline
\begin{tabular}[c]{@{}l@{}}ResBlock \\ (No Shortcut)\end{tabular} & - & 4 $\times$ 2 $\times$ 16. ch \\ \hline
ReLU & - & 4 $\times$ 2 $\times$ 16. ch \\ \hline
Global sum pooling & - & 1 $\times$ 1 $\times$ 16. ch \\ \hline
Concat with input feature  & - & 256+128=384 \\ \hline
Dense & - & 128 \\ \hline
ReLU & - & 128 \\ \hline
Dense & - & 1 \\ \hline

\end{tabular}
\label{tab:tab10}
\end{table}

\subsection{GAAE}
In the GAAE model, the downsampling and upsampling layers are the same as those shown in table \ref{tab:tab1} and \ref{tab:tab2}, respectively.

The Encoder architecture is given in table \ref{tab:tab11}, where we use two dense layers to obtain $z_{x_{u}}$ and $z^{'}_{x_{u}}$ from a global sum pooling layer. For the Decoder, the conditional vector $y_{r}$ or $\hat{y}_{x_{u}}$ is given through the conditional Batch Normaliser (cBN) from the upsampling layer. The classifier network is built upon some dense layer, and the architecture is given in table \ref{tab:tab13}. For the Sample Discriminator, we exactly follow the implementation in table \ref{tab:tab4}. Here, in the table \ref{tab:tab4}, $y$ is the conditional vector, and $h$ is the output from the global sum pooling layer. For the Latent Discriminator, we have use multi dense layers, and the architecture is given in table \ref{tab:tab14}.

The learning rates for both Discriminators are $0.0002$, and for other networks, the learning rate is $0.00005$. We set the number of channels to $16$ for all the experiment carried out with the GAAE.

\subsection{Simple Classifier}

For many classification tasks, we mention a Simple Classifier throughout the paper. The architecture of these classifiers are as in table \ref{tab:tab15}. Here, $c$ is the number of outputs according to the classification categories. The learning rates is used as $0.0001$ for this classifier network.

\begin{table}[ht!]
\centering
\caption{Architecture of the Encoder for the GAAE.}
\begin{tabular}{|l|l|l|}
\hline
\textbf{\begin{tabular}[c]{@{}l@{}}Layer\\ Name\end{tabular}} & \textbf{RS} & \textbf{\begin{tabular}[c]{@{}l@{}}Output\\ Size\end{tabular}} \\ \hline
\begin{tabular}[c]{@{}l@{}}Input \\ Spectrogram\end{tabular} & - & 256 $\times$ 128 $\times$ 1 \\ \hline
ResBlock & D & 128 $\times$ 64 $\times$ 1. ch \\ \hline
Non-local block & - & 128 $\times$ 64 $\times$ 1. ch \\ \hline
ResBlock & - & 64 $\times$ 32 $\times$ 1. ch \\ \hline
ResBlock & D & 32 $\times$ 16 $\times$ 2. ch \\ \hline
ResBlock & D & 16 $\times$ 8 $\times$ 4. ch \\ \hline
ResBlock & D & 8 $\times$ 4 $\times$ 8. ch \\ \hline
ResBlock & D & 4 $\times$ 2 $\times$ 16. ch \\ \hline
\begin{tabular}[c]{@{}l@{}}ResBlock \\ (No Shortcut)\end{tabular} & - & 4 $\times$ 2 $\times$ 16. ch \\ \hline
ReLU & - & 4 $\times$ 2 $\times$ 16. ch \\ \hline
Global sum pooling & - & 1 $\times$ 1 $\times$ 16. ch \\ \hline
Dense ($z_{x_{u}}$), Dense ($z^{'}_{x_{u}}$) & - & 128, 128 \\ \hline

\end{tabular}
\label{tab:tab11}
\end{table}

\begin{table}[h!]
\centering
\caption{Architecture of the Decoder for the GAAE.}
\begin{tabular}{|l|l|l|l|}
\hline
\textbf{\begin{tabular}[c]{@{}l@{}}Layer\\ Name\end{tabular}} & \textbf{RS} & \textbf{SN} & \textbf{\begin{tabular}[c]{@{}l@{}}Output\\ Size\end{tabular}} \\ \hline
Input latent vector & - & - & 128 \\ \hline
Dense & - & - & 4 $\times$ 2 $\times$ 16. ch \\ \hline
ResBlock & U & SN & 8 $\times$ 4 $\times$ 16. ch \\ \hline
ResBlock & U & SN & 16 $\times$ 8 $\times$ 16. ch \\ \hline
ResBlock & U & SN & 32 $\times$ 16 $\times$ 16. ch \\ \hline
ResBlock & U & SN & 64 $\times$ 32 $\times$ 16. ch \\ \hline
ResBlock & U & SN & 128 $\times$ 64 $\times$ 16. ch \\ \hline
Non-local block & - & - & 128 $\times$ 64 $\times$ 16. ch \\ \hline
ResBlock & U & SN & 256 $\times$ 128 $\times$ 1. ch \\ \hline
BN, ReLU & - & - & 256 $\times$ 128 $\times$ 1 \\ \hline
Conv {[}3, 3, 1{]} & - & - & 256 $\times$ 128 $\times$ 1 \\ \hline
Tanh & - & - & 256 $\times$ 128 $\times$ 1 \\ \hline

\end{tabular}
\label{tab:tab12}
\end{table}

\begin{table}[h!]
\centering
\caption{Architecture of the Classifier for the GGAN.}
\begin{tabular}{|l|l|}
\hline
\textbf{\begin{tabular}[c]{@{}l@{}}Layer\\ Name\end{tabular}} & \textbf{\begin{tabular}[c]{@{}l@{}}Output\\ Size\end{tabular}} \\ \hline
Input latent vector & 128 \\ \hline
Dense & 128 \\ \hline
ReLU & 128 \\ \hline
Dense & 10 \\ \hline

\end{tabular}
\label{tab:tab13}
\end{table}

\begin{table}[h!]
\centering
\caption{Architecture of the Latent Discriminator for the GGAN.}
\begin{tabular}{|l|l|}
\hline
\textbf{\begin{tabular}[c]{@{}l@{}}Layer\\ Name\end{tabular}} & \textbf{\begin{tabular}[c]{@{}l@{}}Output\\ Size\end{tabular}} \\ \hline
Input latent vector & 128 \\ \hline
Dense & 128 \\ \hline
ReLU & 128 \\ \hline
Dense & 128 \\ \hline
ReLU & 128 \\ \hline
Dense & 1 \\ \hline

\end{tabular}
\label{tab:tab14}
\end{table}

\begin{table}[h!]
\centering
\caption{Architecture of the Simple Spectrogram Classifier.}
\begin{tabular}{|l|l|}
\hline
\textbf{\begin{tabular}[c]{@{}l@{}}Layer\\ Name\end{tabular}} & \textbf{\begin{tabular}[c]{@{}l@{}}Output\\ Size\end{tabular}} \\ \hline
\begin{tabular}[c]{@{}l@{}}Input \\ Spectrogram\end{tabular} & 256 $\times$ 128 $\times$ 1 \\ \hline
Convolution [3, 3, 32] & 256 $\times$ 128 $\times$ 32 \\ \hline
Maxpool [2, 2] & 128 $\times$ 64 $\times$ 32 \\ \hline
Convolution [3, 3, 64] & 128 $\times$ 64 $\times$ 64 \\ \hline
Maxpool [2, 2] & 64 $\times$ 32 $\times$ 64 \\ \hline
Convolution [3, 3, 128] & 64 $\times$ 32 $\times$ 128 \\ \hline
Maxpool [2, 2] & 32 $\times$ 16 $\times$ 128 \\ \hline
Convolution [3, 3, 256] & 32 $\times$ 16 $\times$ 256 \\ \hline
Maxpool [2, 2] & 16 $\times$ 8 $\times$ 256 \\ \hline
Dense & c \\ \hline

\end{tabular}
\label{tab:tab15}
\end{table}

\begin{IEEEbiography}[{\includegraphics[width=1.1in,height=1.1in,clip]{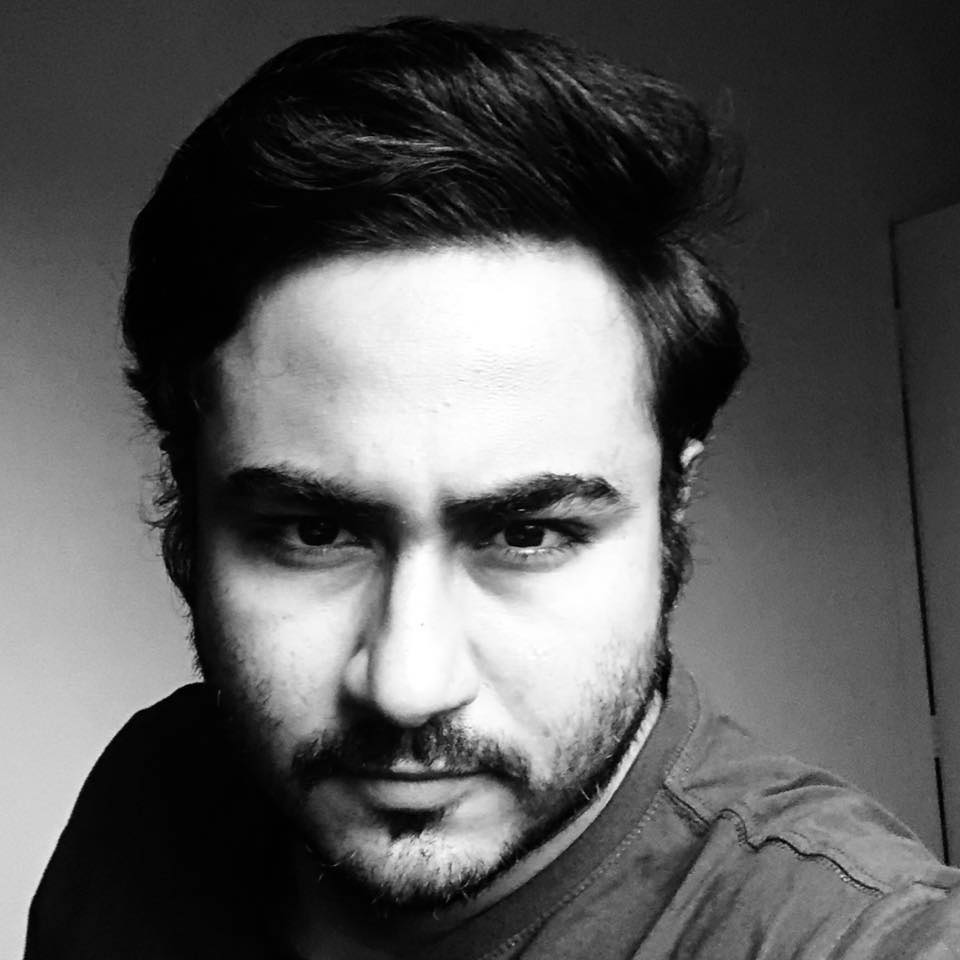}}]{Kazi Nazmul Haque}
is a PhD student at University of Southern Queensland, Australia. He has been working professionally in the field of machine learning for more than five years. Kazi's research work focuses on building machine learning models to solve diverse real-world problems. The current focus of his research work is unsupervised representation learning for 
%BS: no "the"
audio data. He has completed his Master in Information Technology from Jahangirnagar University, Bangladesh. 
\end{IEEEbiography}

\begin{IEEEbiography}[{\includegraphics[width=1.1in,height=1.1in,clip]{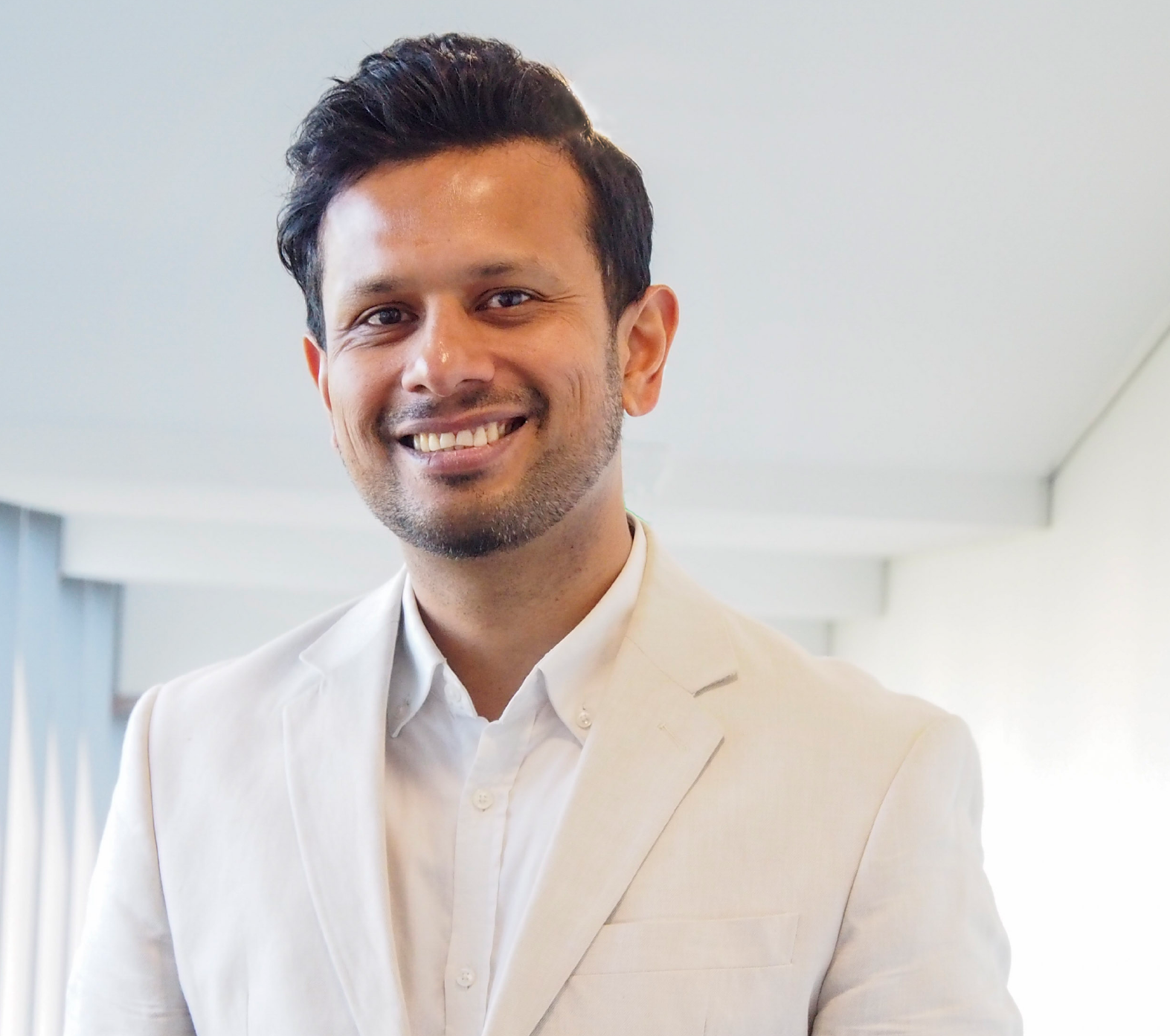}}]{Rajib Rana}
is an experimental computer scientist, Advance Queensland Research Fellow and a Senior Lecturer in the University of Southern Queensland. He is also the Director of 
%BS: 
the 
IoT Health research program at the University of Southern Queensland. He is the recipient of the prestigious Young Tall Poppy QLD Award 2018 as one of 
%BS: 
Queensland's 
most outstanding scientists for achievements in the area of scientific research and communication. Rana's research work aims to capitalise on advancements in technology along with sophisticated information and data processing to better understand disease progression in chronic health conditions and develop predictive algorithms for chronic diseases, such as mental illness and cancer. His current research focus is on Unsupervised Representation Learning. He received his 
%BS: 
B.\,Sc.\ 
degree in Computer Science and Engineering from Khulna University, Bangladesh with 
%BS: 
the 
Prime Minister and 
%BS: 
President's Gold medal for outstanding achievements and 
%BS: 
a Ph.\,D.\ 
in Computer Science and Engineering from the University of New South Wales, Sydney, Australia in 2011. He received his postdoctoral training at Autonomous Systems Laboratory, CSIRO before joining the University of Southern Queensland as Faculty in 2015.
\end{IEEEbiography}

%BS: updated a bit...:
\begin{IEEEbiography}[{\includegraphics[width=1in,height=1.25in,clip,keepaspectratio]{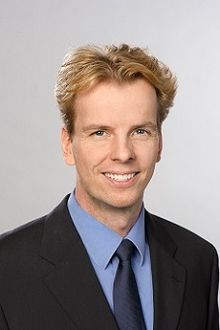}}]{Bj\"{o}rn W.\ Schuller  (M'05-SM'15-F'18)} received his diploma in 1999, his doctoral degree for his study
on Automatic Speech and Emotion Recognition in 2006, and his habilitation and Adjunct Teaching Professorship in the subject area of Signal Processing and Machine Intelligence in 2012, all
in electrical engineering and information technology from TUM in Munich/Germany. He is Professor of Artificial Intelligence in the Department of Computing at the Imperial College London/UK, where he heads GLAM –- the Group on Language, Audio \& Music, Full Professor and head of the Chair of Embedded Intelligence for Health Care and Wellbeing at the University of Augsburg/Germany, and founding CEO/CSO of audEERING. He was previously full professor and head of the Chair of Complex and Intelligent Systems at the University of Passau/Germany. Professor Schuller is Fellow of the IEEE, Golden Core Member of the IEEE Computer Society, Fellow of the ISCA, Senior Member of the ACM, President-emeritus of the Association for the Advancement of Affective Computing (AAAC), and was elected member of the IEEE Speech and Language Processing Technical Committee. He (co-)authored 5 books and more than 900 publications in peer-reviewed books, journals, and conference proceedings leading to more than overall 30\,000 citations (h-index = 82). Schuller is Field Chief Editor of Frontiers in Digital Health, former Editor in Chief of the IEEE Transactions on Affective Computing, and was general chair of ACII 2019, co-Program Chair of Interspeech 2019 and ICMI 2019, repeated Area Chair of ICASSP, next to a multitude of further Associate and Guest Editor roles and functions in Technical and Organisational Committees.
\end{IEEEbiography}

\EOD

\end{document}